%% file: HIG-22-009_temp.tex
\pdfoutput=1
\documentclass[11pt,twoside,a4paper,cmspaper,final,collab]{cms-tdr}
\def\svnVersion{7650f58}\def\svnDate{2024/02/03}\def\cmsCernNoTag{CERN-EP-2023-110}\def\cmsCernDate{\today}\def\cmsMessage{Published in the Journal of High Energy Physics as \href{http://dx.doi.org/10.1007/JHEP01(2024)173}{\doi{10.1007/JHEP01(2024)173}.}}
\begin{document}\cmsNoteHeader{HIG-22-009}

\newlength\cmsTabSkip\setlength{\cmsTabSkip}{1ex}
\newcommand{\bb}{\bbbar}
\newcommand{\mbb}{\ensuremath{m_{\bb}}\xspace}
\newcommand{\qqH}{\ensuremath{\Pq\Pq\PH}\xspace}
\newcommand{\qqZ}{\ensuremath{\PQq\PAQq\to\PZ}\xspace}
\newcommand{\WWZ}{\ensuremath{\PW\PW\to\PZ}\xspace}
\newcommand{\dphibb}{\ensuremath{\Delta\phi_{\bb}}\xspace}
\newcommand{\dphijj}{\ensuremath{\Delta\phi_\mathrm{jj}}\xspace}
\newcommand{\phibb}{\ensuremath{\phi_{\bb}}\xspace}
\newcommand{\phijj}{\ensuremath{\phi_\mathrm{jj}}\xspace}
\newcommand{\mjj}{\ensuremath{m_\mathrm{jj}}\xspace}
\newcommand{\detajj}{\ensuremath{\Delta\eta_\mathrm{jj}}\xspace}
\newcommand{\ggF}{\ensuremath{\Pg\Pg\PH}\xspace}
\newcommand{\VBF}{\ensuremath{\text{VBF}}\xspace}
\newcommand{\ttH}{\ensuremath{\ttbar\PH}\xspace}
\newcommand{\VH}{\ensuremath{\PV\PH}\xspace}
\newcommand{\Hbb}{\ensuremath{\PH\to\bbbar}\xspace}
\newcommand{\Hcc}{\ensuremath{\PH\to\ccbar}\xspace}
\newcommand{\qqHbb}{\ensuremath{\qqH\to\Pq\Pq\bbbar}\xspace}
\newcommand{\Zbb}{\ensuremath{\PZ\to\bbbar}\xspace}
\newcommand{\hbb}{\ensuremath{\PH\bbbar}\xspace}
\newcommand{\zbb}{\ensuremath{\PZ\bbbar}\xspace}
\newcommand{\QCD}{\ensuremath{\text{QCD}}\xspace}
\newcommand{\zjets}{{\PZ}+jets\xspace}
\newcommand{\wjets}{{\PW}+jets\xspace}
\newcommand{\gjets}{photon+jets\xspace}
\newcommand{\HLTTightVBF}{{\texttt{HLT\_Tight}}\xspace}
\newcommand{\HLTLooseVBF}{{\texttt{HLT\_Loose}}\xspace}
\newcommand{\TightVBF}{{\texttt{Tight}}\xspace}
\newcommand{\LooseVBF}{{\texttt{Loose}}\xspace}
\newcommand{\TightVBFA}{{\texttt{Tight 2016}}\xspace}
\newcommand{\LooseVBFA}{{\texttt{Loose 2016}}\xspace}
\newcommand{\TightVBFB}{{\texttt{Tight 2018}}\xspace}
\newcommand{\LooseVBFB}{{\texttt{Loose 2018}}\xspace}

\newcommand{\tabTightAOne}{{\texttt{Tight 1}}\xspace}
\newcommand{\tabTightATwo}{{\texttt{Tight 2}}\xspace}
\newcommand{\tabTightAThree}{{\texttt{Tight 3}}\xspace}

\newcommand{\tabTightBOne}{{\texttt{Tight 1}}\xspace}
\newcommand{\tabTightBTwo}{{\texttt{Tight 2}}\xspace}
\newcommand{\tabTightBThree}{{\texttt{Tight 3}}\xspace}

\newcommand{\tabLooseAOne}{{\texttt{Loose G1}}\xspace}
\newcommand{\tabLooseATwo}{{\texttt{Loose G2}}\xspace}
\newcommand{\tabLooseAThree}{{\texttt{Loose V1}}\xspace}
\newcommand{\tabLooseAFour}{{\texttt{Loose V2}}\xspace}
\newcommand{\tabLooseAFive}{{\texttt{Loose Z1}}\xspace}
\newcommand{\tabLooseASix}{{\texttt{Loose Z2}}\xspace}

\newcommand{\tabLooseBOne}{{\texttt{Loose G1}}\xspace}
\newcommand{\tabLooseBTwo}{{\texttt{Loose G2}}\xspace}
\newcommand{\tabLooseBThree}{{\texttt{Loose V1}}\xspace}
\newcommand{\tabLooseBFour}{{\texttt{Loose V2}}\xspace}
\newcommand{\tabLooseBFive}{{\texttt{Loose Z1}}\xspace}
\newcommand{\tabLooseBSix}{{\texttt{Loose Z2}}\xspace}

\newcommand{\TightAOne}{{\texttt{Tight 2016 1}}\xspace}
\newcommand{\TightATwo}{{\texttt{Tight 2016 2}}\xspace}
\newcommand{\TightAThree}{{\texttt{Tight 2016 3}}\xspace}

\newcommand{\TightBOne}{{\texttt{Tight 2018 1}}\xspace}
\newcommand{\TightBTwo}{{\texttt{Tight 2018 2}}\xspace}
\newcommand{\TightBThree}{{\texttt{Tight 2018 3}}\xspace}

\newcommand{\LooseASix}{{\texttt{Loose 2016 Z2}}\xspace}
\newcommand{\LooseBSix}{{\texttt{Loose 2018 Z2}}\xspace}

\newcommand{\muHbb}{\ensuremath{\mu^\text{incl.}_{\hbb}}\xspace}
\newcommand{\muZ}{\ensuremath{\mu_{\zbb}}\xspace}
\newcommand{\muVBF}{\ensuremath{\mu^{\qqH}_{\hbb}}\xspace}
\newcommand{\muggF}{\ensuremath{\mu^{\ggF}_{\hbb}}\xspace}
\newcommand{\FxFx}{FxFx\xspace}
\newcommand{\DEEPJET} {\textsc{DeepJet}\xspace}
\newcommand{\MINLO} {\textsc{MiNLO}\xspace}
\newcommand{\DEEPCSV} {\textsc{DeepCSV}\xspace}
\newcommand{\CSV} {\textsc{CSV}\xspace}

\cmsNoteHeader{HIG-22-009}
\title{Measurement of the Higgs boson production via vector boson fusion and its decay into bottom quarks in proton-proton collisions at \texorpdfstring{$\sqrt{s}=13\TeV$}{sqrt(s)=13 TeV}}

\author*[cern]{The CMS Collaboration}

\date{\today}

\abstract{
A measurement of the Higgs boson (\PH) production via vector boson fusion (VBF) and its decay into a bottom quark-antiquark pair (\bbbar) is presented using proton-proton collision data recorded by the CMS experiment at $\sqrt{s}=13\TeV$ and corresponding to an integrated luminosity of 90.8\fbinv.
Treating the gluon-gluon fusion process as a background and constraining its rate to the value expected in the standard model (SM) within uncertainties, the signal strength of the VBF process, defined as the ratio of the observed signal rate to that predicted by the SM, is measured to be ${\muVBF=1.01^{+0.55}_{-0.46}}$.
The VBF signal is observed with a significance of 2.4 standard deviations relative to the background prediction, while the expected significance is 2.7 standard deviations.
Considering inclusive Higgs boson production and decay into bottom quarks, the signal strength is measured to be ${\muHbb=0.99^{+0.48}_{-0.41}}$, corresponding to an observed (expected) significance of 2.6 (2.9) standard deviations.
}

\hypersetup{
        pdfauthor={CMS Collaboration},
        pdftitle={Measurement of the Higgs boson production via vector boson fusion and decay into bottom quarks in proton-proton collisions at sqrt(s)=13 TeV},
        pdfsubject={CMS},
        pdfkeywords={CMS, Higgs, VBF, b quarks}
}

\maketitle

\section{Introduction}
\label{sec:intro}
Following the discovery of a scalar boson with a mass near 125\GeV by the ATLAS and CMS Collaborations~\cite{HiggsdiscoveryAtlas, Chatrchyan:2012ufa, Chatrchyan:2013lba} at the CERN LHC, extensive studies of the properties of this particle have been pursued.
A multitude of measurements performed thus far in various production and decay modes support the hypothesis that the discovered particle is the Higgs boson (\PH) described by the standard model (SM) that arises as a consequence of the Brout--Englert--Higgs mechanism of the electroweak (EW) symmetry breaking~\cite{PhysRevLett.13.321, Higgs:1964ia,PhysRevLett.13.508, PhysRevLett.13.585, Higgs:1966ev, Kibble:1967sv}.

The SM Higgs boson with mass of 125\GeV decays most frequently into a bottom quark-antiquark (\bbbar) pair with a branching fraction of about 58\%~\cite{Djouadi:1997yw,LHCHiggsCrossSectionWorkingGroup:2016ypw}.
However, it is challenging to explore this decay mode experimentally.
In the dominant gluon-gluon fusion (\ggF) production mode, the \Hbb signal is overwhelmed by a background consisting of \bbbar pairs produced through the strong interaction, referred to as quantum chromodynamics (QCD) multijet events.
At the LHC, a moderate sensitivity to the \Hbb decay in the \ggF process can be achieved by exploiting Lorentz-boosted production of the Higgs boson~\cite{ATLAS:2021tbi, CMS:2020zge}.
The most promising production mechanism to study the \Hbb decay is the Higgs boson production in association with a leptonically decaying \PW or \PZ boson (\VH).
Although the \VH production cross section is more than an order of magnitude smaller than that of \ggF, leptonic decays of \PW and \PZ bosons provide a handle to reduce backgrounds, thereby making the \Hbb decay accessible for detection.
The \VH production mode has contributed with the largest sensitivity to the observation of the \Hbb decay by the CMS and ATLAS Collaborations.
The ATLAS Collaboration has measured the \Hbb signal yield relative to the SM prediction to be ${\muHbb=1.02\pm 0.12\stat\pm 0.14\syst}$, corresponding to a significance of 6.7 standard deviations ($\sigma$)~\cite{ATLAS:2020fcp} using 139 \fbinv of $\Pp\Pp$ collisions data collected at the center-of-mass energy ($\sqrt{s}$) of 13\TeV.
The measurement by CMS is ${\muHbb =1.04\pm 0.14\stat\pm 0.14\syst}$ and corresponds to a significance of 5.6\,$\sigma$~\cite{CMS:2018nsn} by combining \VH along with other production modes using the data collected at $\sqrt{s} = 7$, 8 and 13\TeV in 2011--2017.
Both the ATLAS and CMS Collaborations have also pursued studies of the Higgs boson production in association with top quark-antiquark pairs, followed by the \Hbb decay.
The rate of this process relative to the SM prediction has been measured to be $0.35^{+0.36}_{-0.34}$ by the ATLAS Collaboration~\cite{ATLAS:2021qou} and $0.33\pm0.26$ by the CMS Collaboration~\cite{CMS-PAS-HIG-19-011}.

This paper considers the vector boson fusion (VBF) production mechanism for the detection of the \Hbb signal.
The VBF production of the Higgs boson has the second-largest cross section at the LHC and attracts particular attention because it involves large momentum transfer and provides a sensitive probe of momentum-dependent anomalous couplings~\cite{Maltoni:2013sma}.

\begin{figure} [htb!]
        \centering
        \includegraphics[width=.6\linewidth]{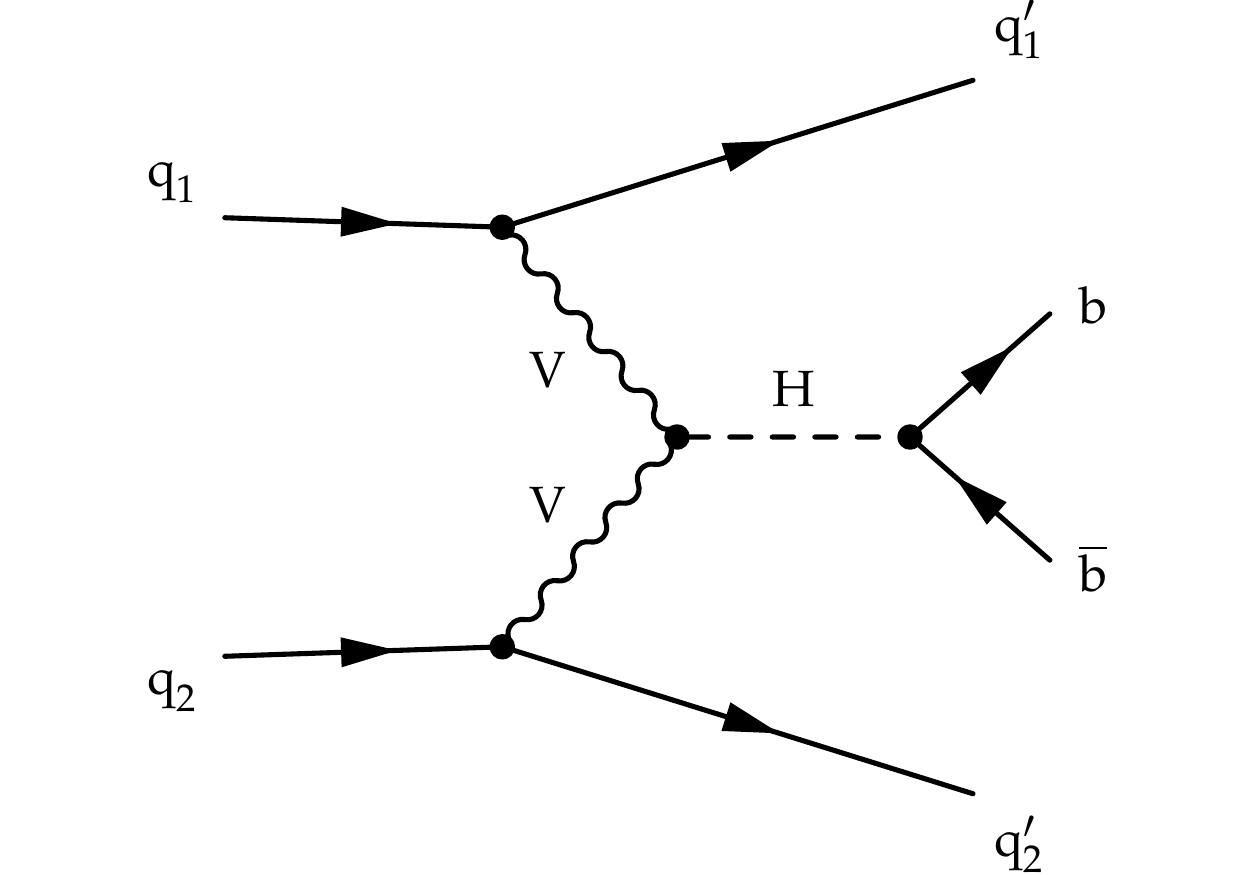}
        \caption{Representative Feynman diagram of the leading order (LO) VBF production of a Higgs boson, followed by its decay to a pair of \PQb quarks.}
        \label{fig:VBF_Hbb}
\end{figure}

The VBF production followed by the \Hbb decay, \qqHbb, gives rise to a four-jet final state as depicted in Fig.~\ref{fig:VBF_Hbb}.
Two of the jets, from the \Hbb decay, typically lie in the central region of the detector.
The other two jets, from the light quarks, are produced mainly in the forward and backward directions relative to the beam line and, consequently, have a large rapidity separation between them, as well as high dijet invariant mass. We refer to the latter as VBF jets.
As the interaction proceeds via the exchange of colorless particles (V = \PW or \PZ), the color connection between the outgoing light-flavor quarks is suppressed, leading to a relatively small amount of hadronic activity in the rapidity interval between the VBF jets and \PQb-tagged jets originating from the Higgs boson decay.
These distinct features allow for the suppression of the large QCD multijet background and the identification of the signal process.

The previous measurement of the \qqHbb process by the CMS Collaboration is based on a data set of proton-proton ($\Pp\Pp$) collisions at $\sqrt{s}=8\TeV$, corresponding to an integrated luminosity of about 20\fbinv~\cite{CMS:2015ebl}.
The signal strength was observed to be ${\muVBF = 2.8^{+1.6}_{-1.4}}$ with a significance of 2.2\,$\sigma$, while the expected significance was 0.8\,$\sigma$.
The ATLAS Collaboration recently reported a measurement of the \qqHbb process using data collected in $\Pp\Pp$ collisions at $\sqrt{s}=13\TeV$, corresponding to an integrated luminosity of 127\fbinv~\cite{ATLAS:2020bhl} and the signal strength was measured to be ${\muVBF=0.95^{+0.38}_{-0.36}}$, corresponding to an observed (expected) significance of 2.6 (2.8)\,$\sigma$.

In this paper, we present a measurement of the \qqHbb process using data collected by CMS at $\sqrt{s}=13\TeV$.
The paper is organized as follows.
In Section~\ref{sec:detector} the main features of the CMS detector are outlined.
Section~\ref{sec:reconstruction} describes the reconstruction of physics objects relevant for this analysis.
Data sets and simulated samples used in the study are presented in Section~\ref{sec:datasets_MC}, and Section~\ref{sec:triggers} describes employed triggers.
Details of the offline analysis, including the event selection and categorization, the improvement of the jet transverse momentum (\pt) resolution by regression techniques, background estimation methods, and signal extraction procedure are discussed in Section~\ref{sec:analysis}.
Section~\ref{sec:syst} details the main systematic uncertainties affecting the analysis.
The final results are discussed in Section~\ref{sec:results} and a brief summary is given in Section~\ref{sec:summary}.
The tabulated results are provided in a HEPData record~\cite{hepdata}.

\section{The CMS detector}
\label{sec:detector}

The central feature of the CMS apparatus is a superconducting solenoid of 6\unit{m} internal diameter, providing a magnetic field of 3.8\unit{T} along the beam direction.
Within the solenoid volume are a silicon pixel and strip tracker, a lead tungstate crystal electromagnetic calorimeter (ECAL), and a brass and scintillator hadron calorimeter (HCAL), each composed of a barrel and two endcap sections.
The ECAL consists of 75\,848 lead tungstate crystals, which provide coverage in pseudorapidity $\abs{\eta} < 1.48 $ in the barrel (EB) region and $1.48 < \abs{\eta} < 3.00$ in the two endcap (EE) regions.
Preshower detectors consisting of two planes of silicon sensors interleaved with a total of 3 radiation lengths of lead are located in front of each endcap detector.
Forward calorimeters extend the $\eta$ coverage provided by the barrel and endcap detectors.
Muons are detected in gas-ionization chambers embedded in the steel flux-return yoke outside the solenoid.

Events of interest are selected using a two-tiered trigger system~\cite{Khachatryan:2016bia}.
The first level (L1), composed of custom hardware processors, uses information from the calorimeters and the muon subdetectors to select events at a rate of around 100\unit{kHz} within a fixed time interval of less than 4\mus.
The second level, known as the high-level trigger (HLT), consists of a farm of processors running a version of the full event reconstruction software optimized for fast processing, and reduces the event rate to around 1\unit{kHz} before data storage~\cite{CMS_HLT}.

A more detailed description of the CMS detector, together with a definition of the coordinate system used and the relevant kinematic variables, can be found in Ref.~\cite{CMSdetector}.

\section{Event reconstruction}
\label{sec:reconstruction}

The reconstruction of the physics objects at CMS is based on the particle-flow (PF) algorithm~\cite{ParticleFlow}, which aims to reconstruct and identify each individual particle (PF candidate) in an event, with an optimized combination of information from the various elements of the CMS detector.

For each event, hadronic jets are clustered from these reconstructed particles using the infrared- and collinear-safe anti-\kt algorithm~\cite{AntiKt, Cacciari:2011ma} with a distance parameter of 0.4.
The jet momentum is determined as the vectorial sum of all particle momenta in the jet, which is typically within 5 to 10\% of the true momentum over the entire range of kinematically allowed \pt and the detector acceptance.
Additional $\Pp\Pp$ interactions within the same or nearby bunch crossings (pileup) increase the track multiplicity and calorimetric energy depositions which, potentially affects the jet momentum.
To mitigate this effect, charged particles identified to be originating from pileup vertices are discarded and an offset correction is applied to account for the remaining contributions.
Jet energy corrections are derived from simulation to match the measured response of the reconstructed jets to that of particle level-jets on average.
In-situ measurements of the momentum balance in dijet, \gjets, \zjets and multijet events are used to account for any residual differences in the jet energy scale between data and simulation~\cite{JECperformance}.
The jet energy resolution amounts typically to 15--20\% at 30\GeV, 10\% at 100\GeV, and 5\% at 1\TeV~\cite{JECperformance}.
Additional selection criteria are applied to remove jets potentially dominated by anomalous contributions from various subdetector components or reconstruction failures.

At the HLT, identification of the jets resulting from the hadronization of \PQb-quarks is performed with the combined secondary vertex algorithm (\CSV)~\cite{Chatrchyan:2012jua,BTV-16-002} and with its improved version using deep machine learning technique (\DEEPCSV)~\cite{BTV-16-002}.
These algorithms combine the information from track impact parameters and secondary vertices identified within a given jet, and provide a continuous discriminator output.
A jet with a {\CSV} or \DEEPCSV discriminant value above a certain threshold is considered to be \PQb-tagged.
The efficiency for tagging \PQb jets and the rate at which light-flavor, gluon, and charm jets are misidentified as \PQb jets depend on the chosen threshold, as well as on the jet \pt and $\eta$.
The thresholds are chosen to ensure an online selection of \PQb jets with an efficiency of 60--70\%, corresponding to a misidentification rate of 2--3\%.
Identification of \PQb jets in the offline analysis is performed with the \DEEPJET tagger~\cite{Bols:2020bkb, CMS-DP-2018-058}, which combines features of the individual jet constituents, properties of the reconstructed secondary vertices, and the global jet variables into a multiclass deep neural network (DNN) targeting three separate classes for the \PQb jets: i) jet with one \PB hadron, ii) jet with two or more \PB hadrons, and iii) jet with semileptonic decays of \PB hadrons.
The sum of three \PQb tag class DNN outputs quantifies the consistency of a jet with the \PQb jet hypothesis.
A jet is considered to be \PQb-tagged in the offline selection, if this DNN output, referred to as the \PQb tag score, exceeds a certain threshold.
These thresholds are chosen such that the \PQb jet selection efficiency is 70--80\% depending on data-taking periods, while the probability of light and gluon jets being misidentified as \PQb jets is about 1\% and the probability of charm jets being misidentified as \PQb jets is about 10\%.
To minimize the discrepancy in the \PQb tagging performance between data and simulation both at the HLT and in the offline analysis, separate jet flavor dependent calibration factors are applied to each jet in simulated events as a function of jet \pt, $\eta$, and the \PQb tag discriminator score.
These corrections have been derived using control samples of the semileptonic \ttbar decays, \zjets events, and the inclusive sample of QCD multijet events~\cite{BTV-16-002}.
Determination of corrections at the HLT is facilitated by the control and prescaled single- and multi-jet triggers.

\section{Data set and simulated samples}
\label{sec:datasets_MC}

This analysis uses data collected by the CMS experiment in $\Pp\Pp$ collisions at $\sqrt{s} = 13\TeV$ corresponding to an integrated luminosity of 90.8\fbinv.
The total analyzed data volume comprises two sets recorded under different experimental conditions.
The first set was collected in 2016 and corresponds to an integrated luminosity of 36.3\fbinv.
The second set was collected in 2018, when the LHC delivered $\Pp\Pp$ collisions at higher instantaneous luminosity compared to 2016, and corresponds to an integrated luminosity of 54.5\fbinv.
No trigger path suitable for this analysis was available during the data-taking periods in 2017.

The analysis relies on simulated Monte Carlo (MC) samples for the estimation of the signal acceptance and efficiency and determination of the trigger efficiency.
Additionally, the contributions from various subdominant background processes are determined from simulation.

The \qqHbb signal is generated using \POWHEG2.0~\cite{Nason:2004rx,Frixione:2007vw,Alioli:2010xd} at next-to-leading order (NLO) precision in the QCD coupling constant \alpS~\cite{Nason:2009ai}.
A dipole parton shower model~\cite{Cabouat_2018} incorporated in \PYTHIA8.212~\cite{Sjostrand:2014zea} is used for emulating of the initial-final state color flow that takes into account the color connection between the incoming and the outgoing partons.
An alternative \qqHbb sample is also prepared using the \POWHEG matrix element generator interfaced with \HERWIG7 v-7.1\cite{Bellm:2015jjp} for fragmentation and hadronization.
This sample is used only to assess the systematic uncertainty related to the choice of the showering and hadronization model.
The \ggF production of the Higgs boson with at least two accompanying jets has a nonnegligible contribution to the kinematic phase space considered in this analysis.
The \ggF process is generated using the \MINLO event generator at next-to-NLO (NNLO)~\cite{Luisoni:2013cuh, Hamilton:2012rf} precision in \alpS~\cite{Bagnaschi:2011tu}, including finite top quark mass effects.
An NLO process that may include loops needs to be hadronized with a parton shower model beyond LO.
As a result, jet matching must take into account additional hard radiative jets.
This is implemented, for this analysis, by employing the \FxFx scheme~\cite{fxfx}.
Contributions from the weak gauge boson associated (\VH) and top-quark-pair associated (\ttH) production of the Higgs boson are found to be negligible.

The dominant continuum background for this analysis is QCD multijet production.
To assess properties of these events and validate the analysis strategy, QCD multijet events are generated with \MGvATNLO~\cite{Alwall:2014hca} at leading order (LO) precision in \alpS.
The matrix element is matched to the parton showers generated by \PYTHIA using the MLM prescription~\cite{MLM_matching}.
Generated samples of QCD multijet production are also used to derive corrections applied to simulated samples to account for differences in the trigger efficiency between data and simulation.

The dominant resonant background is the inclusive production of \PZ bosons (\zjets).
About 70\% of the time, \PZ bosons decay via a quark-antiquark pair of the same flavor, including a 15\% branching fraction into a pair of bottom quarks.
Hence, the main component of the resonant background in the event sample corresponds to the \Zbb decay mode, although there are contributions from charm and light-flavor quark jets being misidentified as \PQb jets.
There are two different mechanisms of \zjets production: via quark-antiquark annihilation (\qqZ) and fusion of \PW bosons (\WWZ).
Though the latter has a much lower rate, the event topology is the same as for the signal.
\PW bosons can be produced in association with jets in similar ways.
The inclusive \PW production is generated using \MGvATNLO at LO precision in \alpS.
The same generator is used to produce mutually exclusive \qqZ and \WWZ samples.
The \qqZ process is simulated considering Feynman diagrams involving direct electroweak coupling of the \PZ boson to quark-antiquark pairs.
In the generation of the \WWZ process, only the Feynman diagrams that contain the $\PW\PW\PZ$ triple gauge boson coupling are considered in the computation of the matrix element.
As in the case of the \qqHbb signal, the \WWZ process is simulated using a dipole parton shower model implemented in \PYTHIA.

For the \zjets samples, correction factors have been applied to match the generator-level \pt distributions with analytic predictions available with the highest order accuracy in the perturbative expansion~\cite{Lindert:2017olm}.
Two individual correction factors are applied to the \qqZ process for each \pt bin; first to emulate the spectrum predicted by perturbative QCD with NNLO accuracy and then further reweighting is done to incorporate the higher-order EW effects.
For the \WWZ mechanism, only higher-order EW correction factors are applied.

Other important background contributions in the signal region arise from inclusive single top quark ($\PQt/\PAQt$+$\PX$) and top quark pair ($\ttbar$+$\PX$) production.
These are modeled by \POWHEG~\cite{Alioli:2011as, Alioli:2009je} at the NLO QCD precision.
For the \ttbar process, all possible combinations of the decay modes of the two \PW bosons from the top quark and antiquark decays are considered leading to hadronic, leptonic, and semileptonic final states.

        {
                \tolerance=2000
                All the simulated samples except the \HERWIG signal sample are interfaced with \PYTHIA8.212 for parton showering and fragmentation with the standard \pt-ordered parton shower scheme.
                The underlying event is modeled with \PYTHIA, using the CP5 tune~\cite{Sirunyan:2019dfx} for both of the years.
                The parton distribution functions (PDFs) are taken from the sets of NNPDF3.0~\cite{Ball:2014uwa} and NNPDF3.1~\cite{Ball:2017} for the 2016 and 2018 samples respectively.
                The response of the CMS detector is modeled using the \GEANTfour~\cite{Agostinelli:2002hh} package.
                The event reconstruction is performed with the same algorithms as are used for the data.
                Additional pileup interactions in each bunch crossing are generated with \PYTHIA and added to the simulated samples following a Poisson distribution with the mean value determined in the data.
                The simulated events are weighted such that the pileup distribution in the simulation matches the one observed in the data.
                \par
        }

\section{Triggers}
\label{sec:triggers}

Events are selected with dedicated L1 and HLT selections optimized separately for 2016 and 2018.
At the L1 stage, events are required to have at least three jets with \pt above certain thresholds that were optimized according to the instantaneous luminosity.
The \pt thresholds of 90, 76, and 64\GeV (100, 80, and 70\GeV) were imposed for 2016 (2018).
The presence of a fourth jet is not required at the L1 stage.

An event is accepted by the HLT if it contains at least four jets reconstructed online with the PF algorithm.
Jets are required to have \pt greater than 92, 76, 64, and 15 (105, 88, 76, and 15)\GeV in 2016 (2018).
Two complementary online requirements (HLT paths), as explained below, are also implemented to select events in each of the two sets.
In the following, we refer to these HLT paths as \HLTTightVBF and \HLTLooseVBF.

The \HLTTightVBF path selects events with at least one \PQb-tagged jet among six leading jets according to \pt.
The \PQb tagging is performed with the \CSV (\DEEPCSV) algorithm in 2016 (2018).
The working point chosen for the online \PQb tagging in the \HLTTightVBF path corresponds to \PQb jet efficiency of roughly 60 to 70\% and a misidentification rate for light-flavor quark and gluon jets of 3 to 4\%.
In addition to the \PQb-tagging selection there is also a VBF-selection implemented at the HLT path, which considers only four \pt leading jets.
Among these four jets, the one with the highest {\CSV} (\DEEPCSV) discriminant is considered to be a jet from the Higgs boson decay, and out of the three remaining jets, a jet pair with the largest $\detajj$ is chosen as the pair of VBF-tagged jets.
The last remaining jet is considered to be the other jet from the decay of the Higgs boson.
The \HLTTightVBF path imposes very stringent conditions on the VBF-tagged jets: they must be separated in pseudorapidity by $\detajj>4.1$ (3.5) and have invariant mass $\mjj>500$ (460)\GeV in 2016 (2018).
For the other two jets, which are assigned to the \Hbb decay, the separation in azimuthal angle is required to be $\dphibb<1.6$ (1.9) for 2016 (2018).

In contrast to the \HLTTightVBF path, the \HLTLooseVBF path selects events with at least two \PQb-tagged jets but imposes comparatively lenient requirements on the VBF-tagged jets: $\mjj>240$ (200)\GeV and $\detajj>2.3$ (1.5) for 2016 (2018); while $\dphibb<2.1$ (2.8) in 2016 (2018) is required for \PQb jets.
Requirements imposed by the \HLTTightVBF and \HLTLooseVBF paths in the 2016 and 2018 data-taking periods are summarized in Table~\ref{tab:Tab_evt_Sel}.

The efficiency of the trigger selection is measured with the tag-and-probe method~\cite{CMS:2011aa} using events selected with control triggers.
Trigger scale factors, representing the ratio of the trigger efficiency in data to that in simulated events, are measured as a function of jet \pt and $\abs{\eta}$ for each leg of the four-jet requirement using control and prescaled single-jet triggers.
For the online \PQb tagging, the scale factors are also derived as a function of the \DEEPJET discriminants of the offline tagged jets employing auxiliary and prescaled four-jet triggers.
The applied corrections range from 2 to 7\% for each leg of the four-jet requirement and from 3 to 10\% per jet for the online \PQb tag requirement.

The absolute trigger efficiencies, computed for the inclusive simulated signal samples, are 3.1 (2.3) and 3.5 (2.5)\% in 2016 (2018), for the \HLTTightVBF and \HLTLooseVBF paths, respectively.
The inefficiencies of the trigger paths are mainly driven by the relatively high \pt thresholds imposed on the three leading jets.

\section{Analysis procedures}
\label{sec:analysis}

Events recorded online by the HLT paths described in Section~\ref{sec:triggers}, are reconstructed with improved information related to the detector conditions and calibrations.

\subsection{Event selection}
\label{sec:selection}

An event is discarded if it contains any isolated muon or electron identified through selection criteria that correspond to a selection efficiency of 95\% and a misidentification rate of 1--2\%.
This requirement suppresses contributions from \ttbar and $\PQt/\PAQt$ events with leptonic decays of the \PW bosons.
The event is selected if it contains four jets with \pt greater than 95, 80, 65, and 30\GeV (110, 90, 80, and 30\GeV) in 2016 (2018). Reconstructed physics objects are matched to the corresponding trigger objects.
At least two jets are required to be \PQb tagged with the \DEEPJET algorithm as described in Section~\ref{sec:reconstruction}.
Out of the four \pt-leading jets, the two jets with the highest \DEEPJET \PQb tag scores are used to reconstruct the \Hbb decay candidate.
After the Higgs boson candidate is selected, the two remaining jets are considered as VBF-tagged jet candidates.
The jet with $\pt < 50\GeV$ must also pass an identification criterion designed to reduce the number of selected jets originating from pileup interactions~\cite{CMS:2020ebo}.

Depending on the data-taking periods and the trigger they pass, events are split into four nonoverlapping samples that are analyzed independently.
Events from 2016 (2018) that pass the \HLTTightVBF path are assigned to the \TightVBFA (\TightVBFB) sample; events failing the \HLTTightVBF path but pass the \HLTLooseVBF path comprise the \LooseVBFA (\LooseVBFB) sample.
The selection criteria applied to the VBF- and \PQb-tagged jets in each sample are:
\begin{itemize}
        \item \TightVBFA: $\dphibb<1.6$, $\detajj>4.2$, $\mjj>500\GeV$;
        \item \LooseVBFA: $\dphibb<2.1$, $\detajj>2.5$, $\mjj>250\GeV$;
        \item \TightVBFB: $\dphibb<1.6$, $\detajj>3.8$, $\mjj>500\GeV$; and
        \item \LooseVBFB: $\dphibb<2.1$, $\detajj>2.5$, $\mjj>250\GeV$.
\end{itemize}

The lower section of Table~\ref{tab:Tab_evt_Sel} outlines the selection criteria used in the offline analysis.

\begin{table} [htb!]
        \topcaption{The HLT and offline selection requirements in the four analyzed samples.}
        \label{tab:Tab_evt_Sel}
        \centering
        \begin{tabular}{cccccc}
                Level                     & Requirements                 & \multicolumn{2}{c}{2016 (36.3\fbinv)}   & \multicolumn{2}{c}{2018 (54.5\fbinv)}                                 \\
                \hline
                                          &                              & \TightVBF                               & \LooseVBF                               & \TightVBF    & \LooseVBF    \\[\cmsTabSkip]
                \multirow{4}{*} {HLT}     & \pt thresholds               & \multicolumn{2}{c}{92, 76, 64, 15\GeV}  & \multicolumn{2}{c}{105, 88, 76, 15\GeV}                               \\
                                          & Number of \PQb tags          & $\geq$1                                 & $\geq$2                                 & $\geq$1      & $\geq$2      \\
                                          & \dphibb                      & $<$1.6                                  & $<$2.1                                  & $<$1.9       & $<$2.8       \\
                                          & \detajj                      & $>$4.1                                  & $>$2.3                                  & $>$3.5       & $>$1.5       \\
                                          & \mjj                         & $>$ 500\GeV                             & $>$240\GeV                              & $>$460\GeV   & $>$200\GeV   \\[\cmsTabSkip]
                \hline
                \multirow{6}{*} {Offline} & \pt thresholds               & \multicolumn{2}{c}{95, 80, 65, 30\GeV } & \multicolumn{2}{c}{110, 90, 80, 30\GeV}                               \\
                                          & Jet $\abs{\eta}<4.7$         & $\checkmark$                            & $\checkmark$                            & $\checkmark$ & $\checkmark$ \\
                                          & Lepton veto                  & $\checkmark$                            & $\checkmark$                            & $\checkmark$ & $\checkmark$ \\
                                          & Number of \PQb tags $\geq$ 2 & $\checkmark$                            & $\checkmark$                            & $\checkmark$ & $\checkmark$ \\
                                          & \PQb jet $\abs{\eta}< 2.4$   & $\checkmark$                            & $\checkmark$                            & $\checkmark$ & $\checkmark$ \\
                                          & \dphibb                      & $<$1.6                                  & $<$2.1                                  & $<$1.6       & $<$2.1       \\
                                          & \detajj                      & $>$4.2                                  & $>$2.5                                  & $>$3.8       & $>$2.5       \\
                                          & \mjj                         & $>$500\GeV                              & $>$250\GeV                              & $>$500\GeV   & $>$250\GeV
        \end{tabular}
\end{table}

\subsection{Regression of \texorpdfstring{\PQb}{b} jet energy}
\label{sec:bregression}

\begin{figure} [htb!]
        \centering
        \includegraphics[width=.48\linewidth]{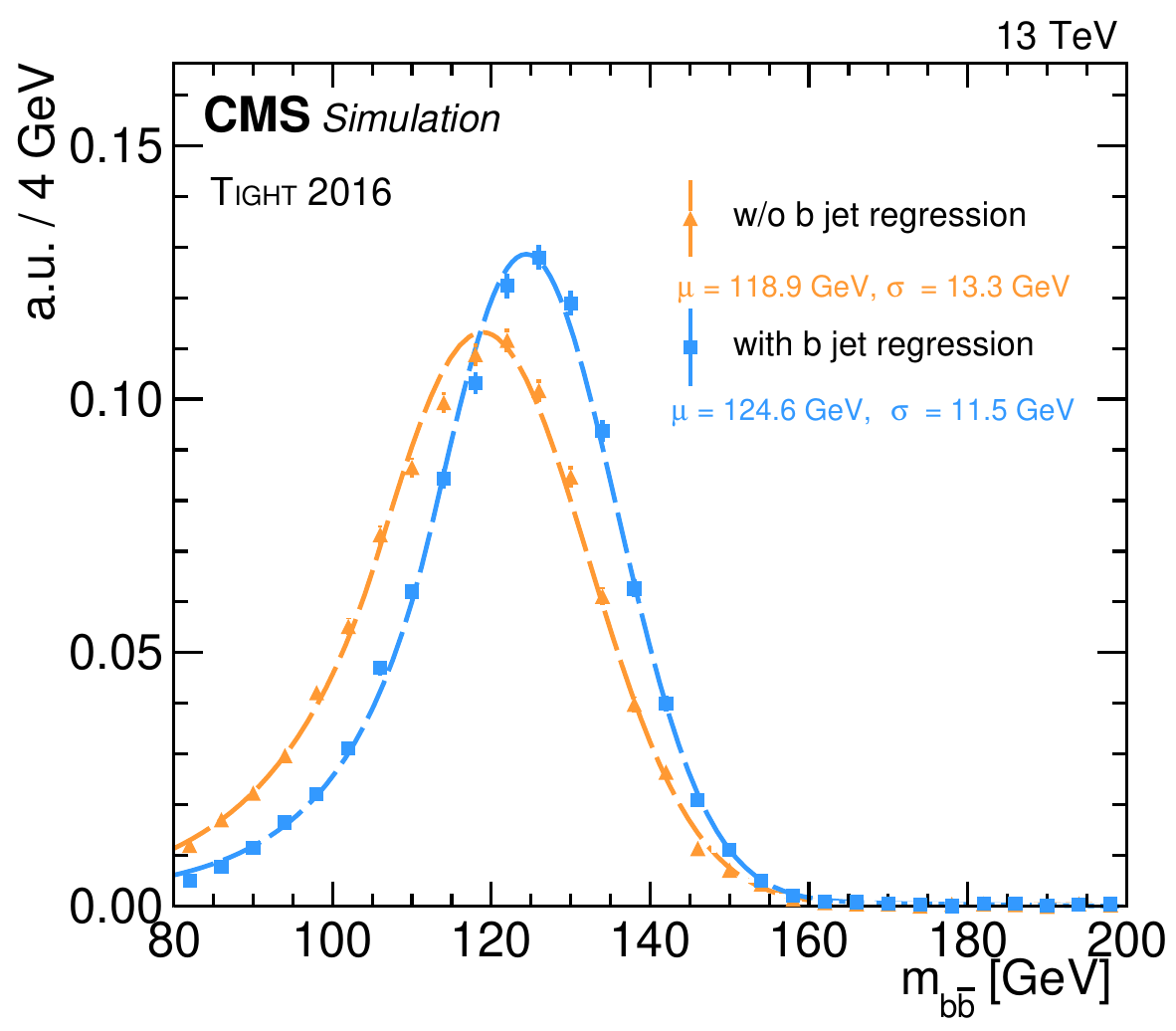}
        \includegraphics[width=.48\linewidth]{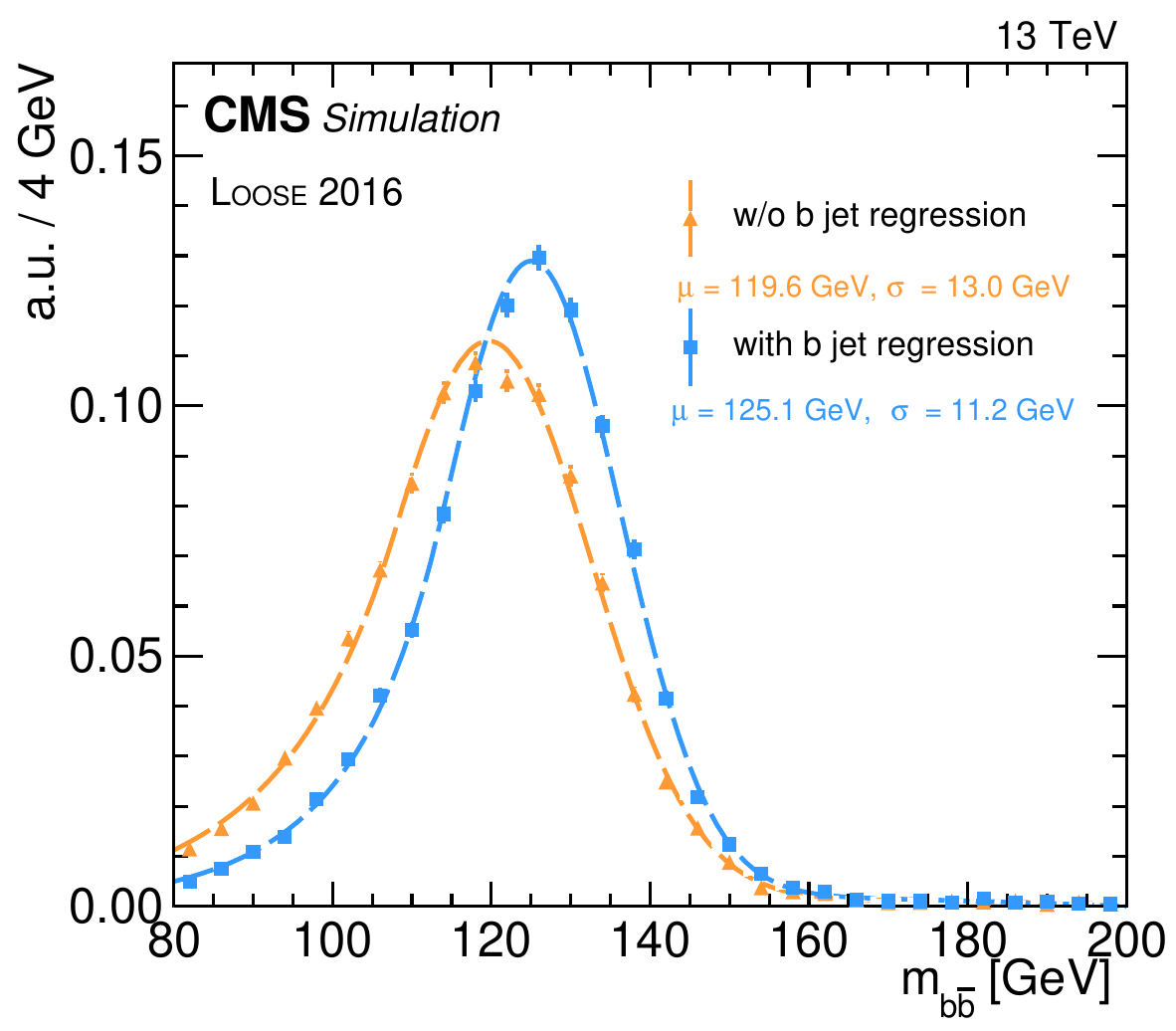}
        \caption{
                The invariant mass \mbb of the \PQb jet pair in simulated \qqHbb events before (orange dashed line) and after (blue dashed line) the application of the \PQb jet energy regression in the \TightVBFA (left) and \LooseVBFA (right) samples.
                A one-sided Crystal Ball function~\cite{CrystalBallRef} is used to fit the distributions, where $\mu$ and $\sigma$ are the peak position and half width of the core part of the CB function, respectively.
        }
        \label{fig:vbf_reg}
\end{figure}

The energy of a \PQb-tagged jet is likely to be underestimated often because of the unmeasured neutrino produced in the semileptonic decay of a \PQb hadron.
This is partially remedied by the energy regression of a \PQb jet~\cite{Sirunyan:2019wwa} that improves the mass resolution of the Higgs boson candidate constructed from the two \PQb-tagged jets.
The regression is based on a DNN trained on simulated \PQb jets from \ttbar events.
The algorithm's input features include properties of the secondary vertices and tracks associated with the jet, jet constituents, soft muons and electrons from the semileptonic decays of \PQb hadrons, as well as additional variables related to the jet energy.
After the application of the \PQb jet energy regression, the jet energy scale and resolution of \PQb jets are corrected in simulated events to match the jet energy scale and resolution observed in data.
Corrections are devised using events where a \PQb-tagged jet recoils against a \PZ boson that decays into leptons.
The \PZ boson \pt is reconstructed with high precision due to the excellent resolution of the lepton momentum.
As the \pt of the \PZ boson is balanced with the jet \pt, the ratio of the momentum of the reconstructed jet to the \PZ boson momentum is used to measure the jet energy scale and resolution in the data and simulated samples and derive corresponding corrections to the energy scale and resolution of \PQb jets in the simulated events.
The correction factors constitute about 0.5--2\% (1--4\%) of the jet \pt for the resolution (scale) of the regressed jets.
As an illustration, Fig.~\ref{fig:vbf_reg} shows the effect of \PQb jet energy regression on the invariant mass \mbb of the reconstructed \PQb jet pair in simulated \qqHbb signal events for the \TightVBFA and \LooseVBFA samples.
Distributions of \mbb are obtained after applying dedicated corrections to the energy scale and resolution of \PQb jets in simulated events.
The \PQb jet energy regression improves the \mbb resolution and the central value by approximately 10 to 15\%.

\subsection{Event categorization}
\label{sec:mva}

To discriminate the signal from the major background sources, a multivariate analysis (MVA) technique is applied; the key properties of the \qqHbb signal are combined in a boosted decision tree (BDT) algorithm implemented with the \textsc{TMVA} package~\cite{Speckmayer:2010zz}.
The BDT response is then used to separate the signal from the background processes.
Two different MVA strategies are applied in the \TightVBF and \LooseVBF samples.

The \TightVBFA and \TightVBFB samples are characterized by lower background yields and higher signal-to-background ratios.
In particular, the contributions from the resonant \Zbb background and \ggF production are significantly suppressed by more stringent requirements imposed on the VBF-tagged jets.
The largest background in these event classes originates from the QCD multijet production.
Two separate binary classifiers are trained to discriminate the \qqHbb signal from the QCD multijet background for each individual year.
The following variables are used as inputs to the BDT:
\begin{itemize}
        \item Properties of the VBF-tagged jets: invariant mass (\mjj), difference in the pseudorapidity (\detajj), and difference in the azimuthal angle (\dphijj);
        \item The minimal opening angle between the momentum vector of any of the two VBF-tagged jets and the momentum vector of the dijet system composed by the VBF-tagged jet pair ($\alpha_\mathrm{jj}$);
        \item The quark-gluon discriminator~\cite{CMS:2013kfa,JetsInRun2} score of the two VBF-tagged jets.
              This is a likelihood discriminator constructed using the information on the charged and the neutral constituents of a given jet and the jet shape variables to distinguish between the jets originating from quarks and from gluons;
        \item The \DEEPJET \PQb-tagging scores of the two selected \PQb-tagged jets assigned to the \Hbb decay;
        \item The ratio of the magnitude of the transverse momentum vector of the selected four-jet system to the scalar \pt-sum of the four selected jets ($\abs{\sum \ptvec}/\sum \pt$);
        \item The longitudinal component of the momentum vector of the selected four-jet system ($\sum p_{z}$);
        \item The difference in the azimuthal angle between the Higgs boson candidate and the dijet system composed of the VBF-tagged jets ($\abs{\phibb-\phijj}$); and
        \item The multiplicity and \pt sum of extra jets with $\pt>30\GeV$ and $\abs{\eta}<2.4$, excluding the selected \PQb-tagged and VBF-tagged jets.
\end{itemize}

\begin{figure} [htb!]
        \centering
        \includegraphics[width=.49\linewidth]{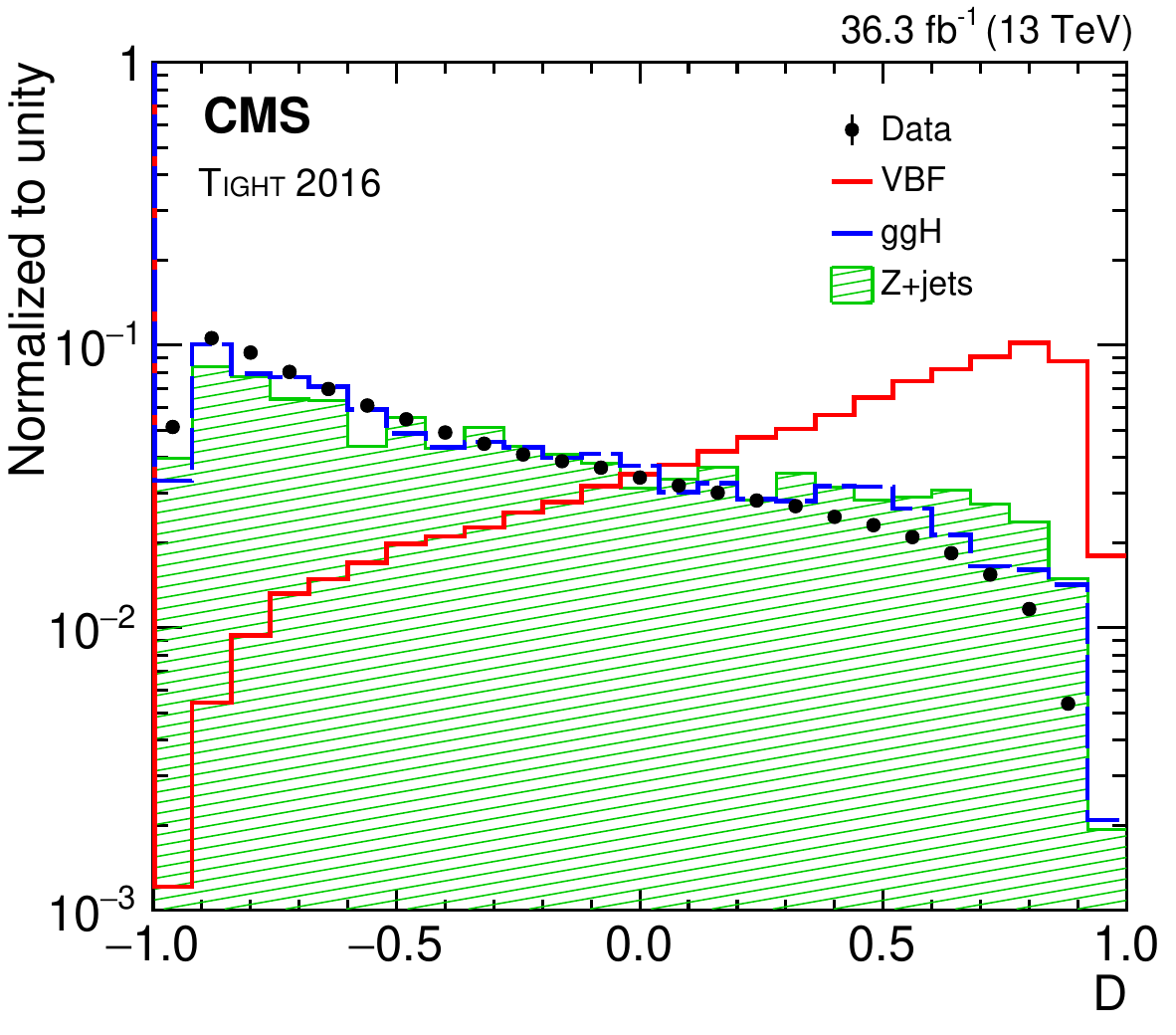}
        \includegraphics[width=.49\linewidth]{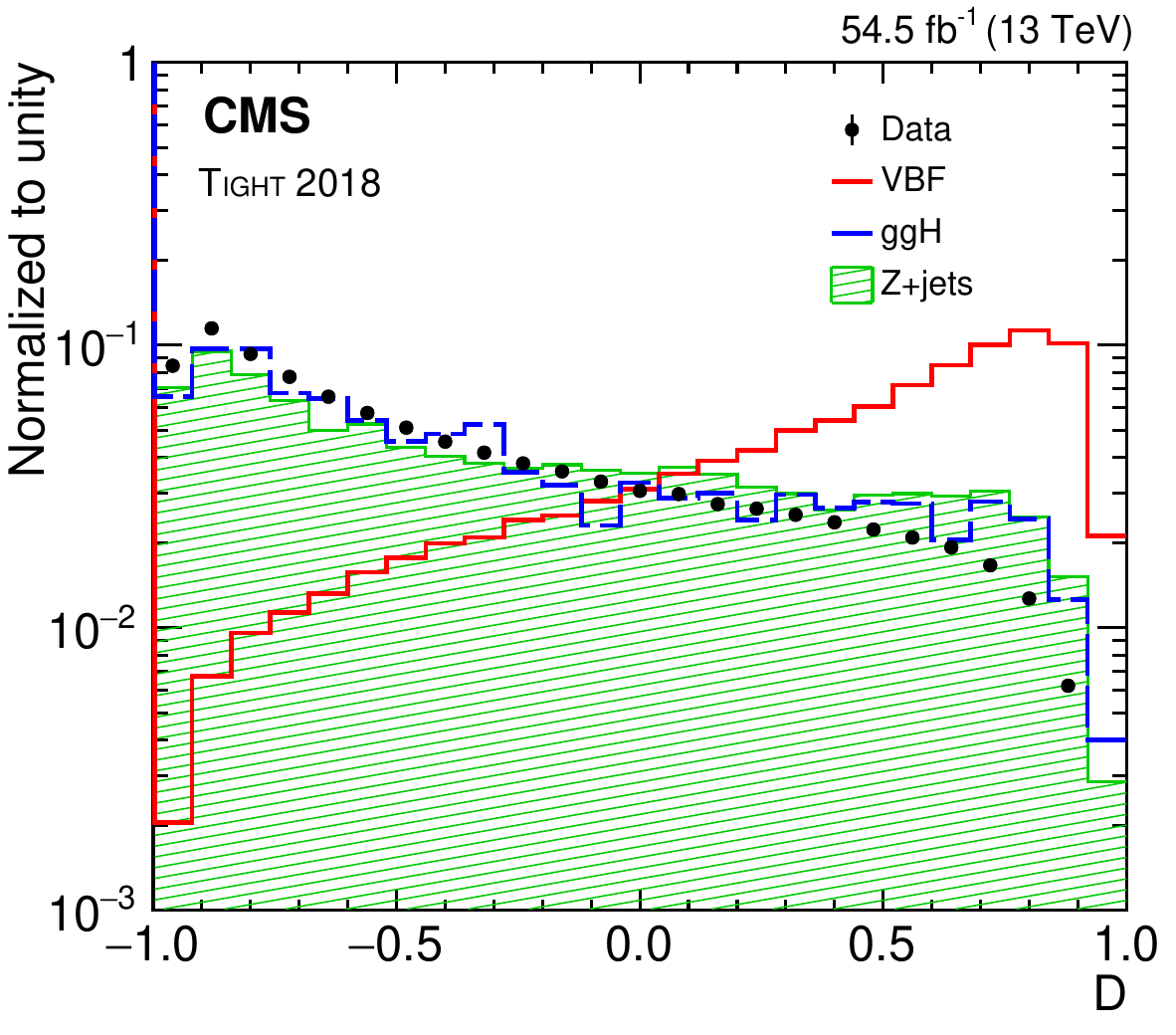}
        \caption{
                The unit normalized distributions of the VBF BDT outputs in data and simulated samples in the \TightVBFA (left) and \TightVBFB (right) analysis samples.
                Data events (points), dominated by the QCD multijet background, are compared to the VBF (red solid line), \ggF (blue dashed line), and \zjets (green hatched area) processes.
        }
        \label{fig:mva_tight}
\end{figure}

The training is performed using half of the simulated VBF events and 2.5\% of the data sample after event selection as a proxy for the dominant QCD multijet background, the rest of the simulated VBF sample as well as another 2.5\% of the data is used for the validation of the training.
Dedicated studies were performed to assess the impact of using data events in the BDT training.
In the first study, several other disjoint data subsets corresponding to 5\% of the total sample were employed in the training and validation of the training.
In the second study, the subset of data used in the training was removed from the data set used for the extraction of signal.
The impact of these alternative analysis strategies on the measured signal rate and its uncertainty was found to be below 2\%.
Figure~\ref{fig:mva_tight} shows the distributions of the BDT output score ($D$) normalized to unity for the VBF, \ggF, and \Zbb processes and the data as an approximation to the QCD multijet background in the \TightVBFA and \TightVBFB samples.

\begin{figure} [htb!]
        \centering
        \includegraphics[width=.47\linewidth]{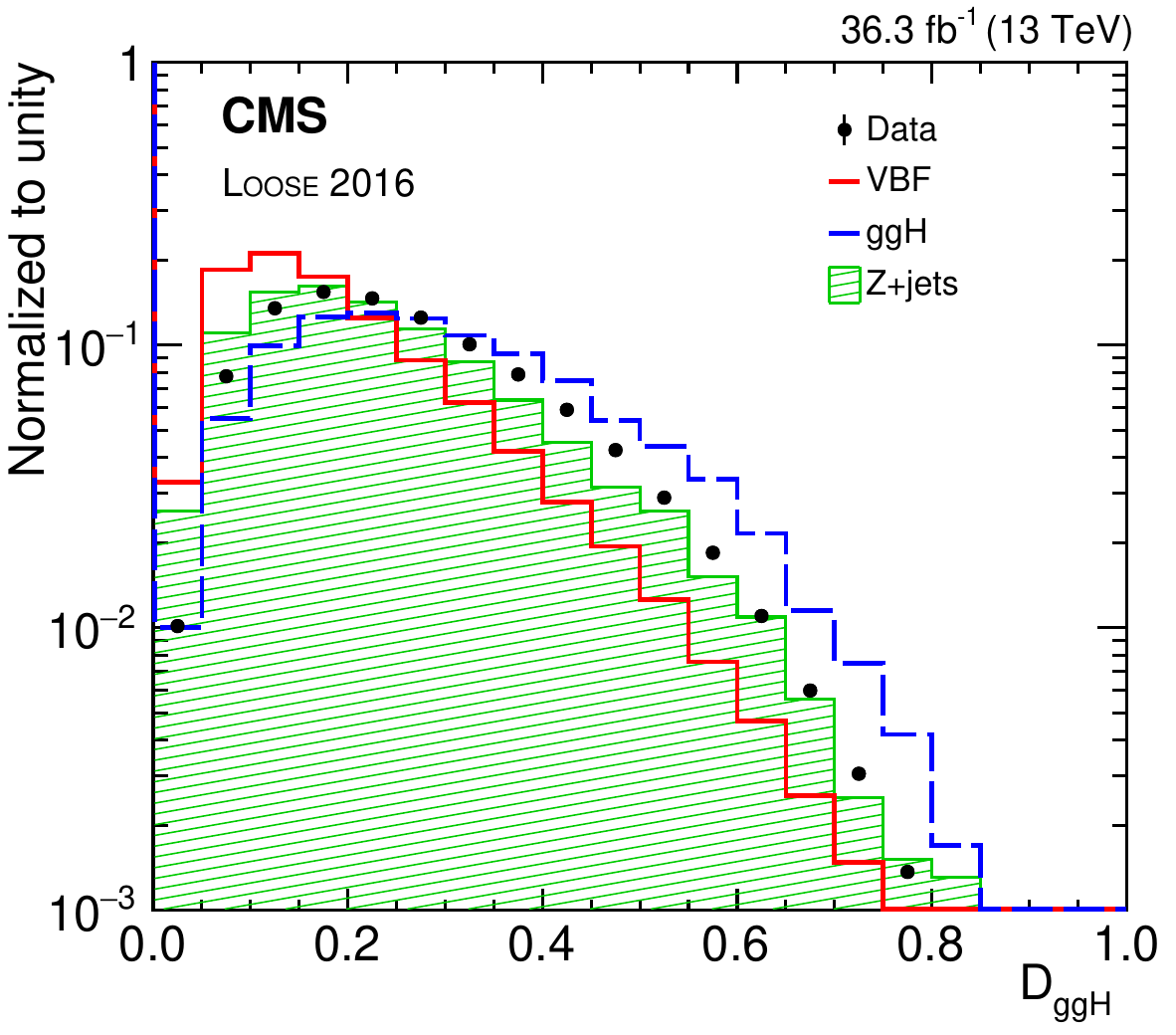}
        \includegraphics[width=.47\linewidth]{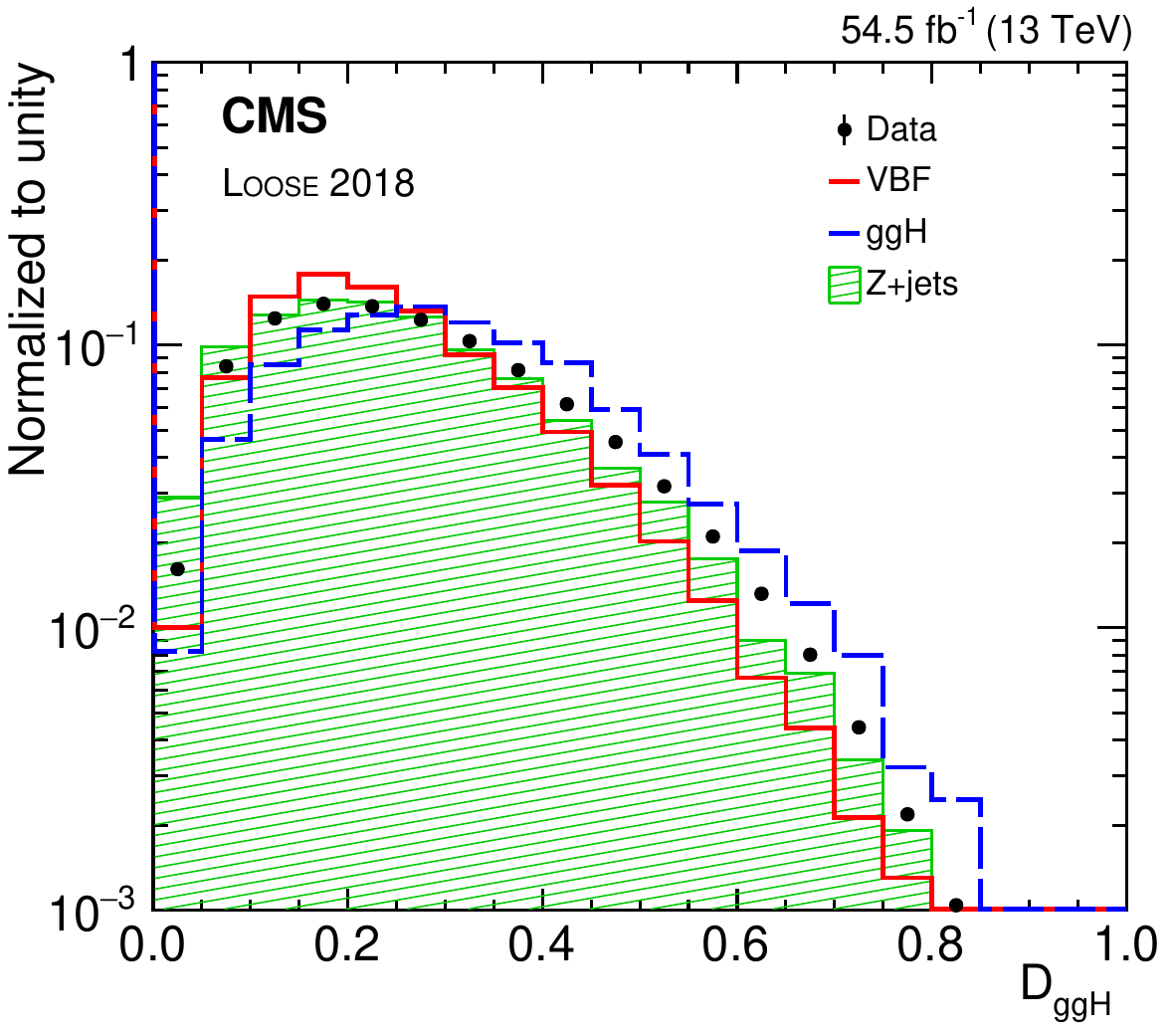}\\
        \includegraphics[width=.47\linewidth]{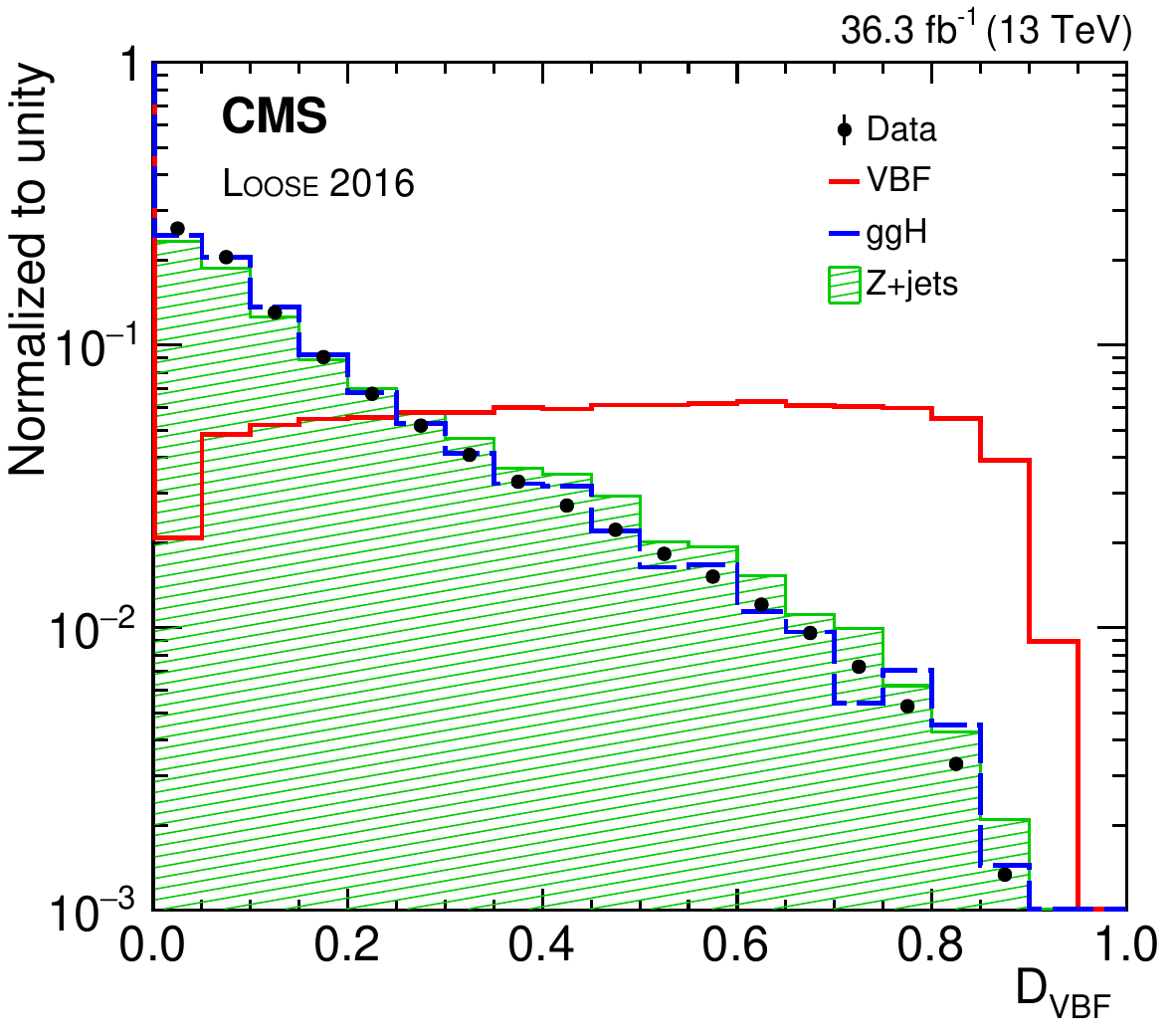}
        \includegraphics[width=.47\linewidth]{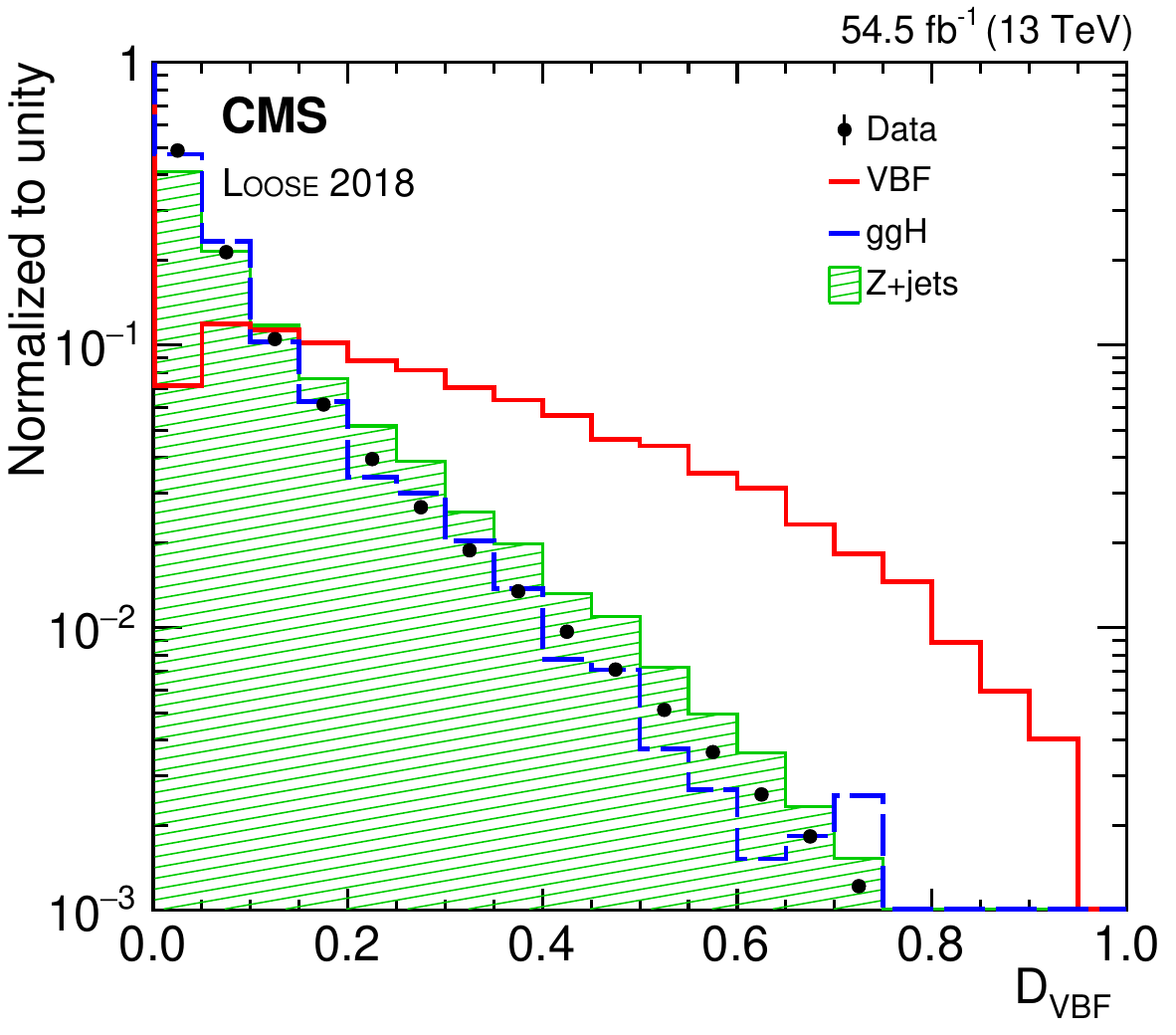}\\
        \includegraphics[width=.47\linewidth]{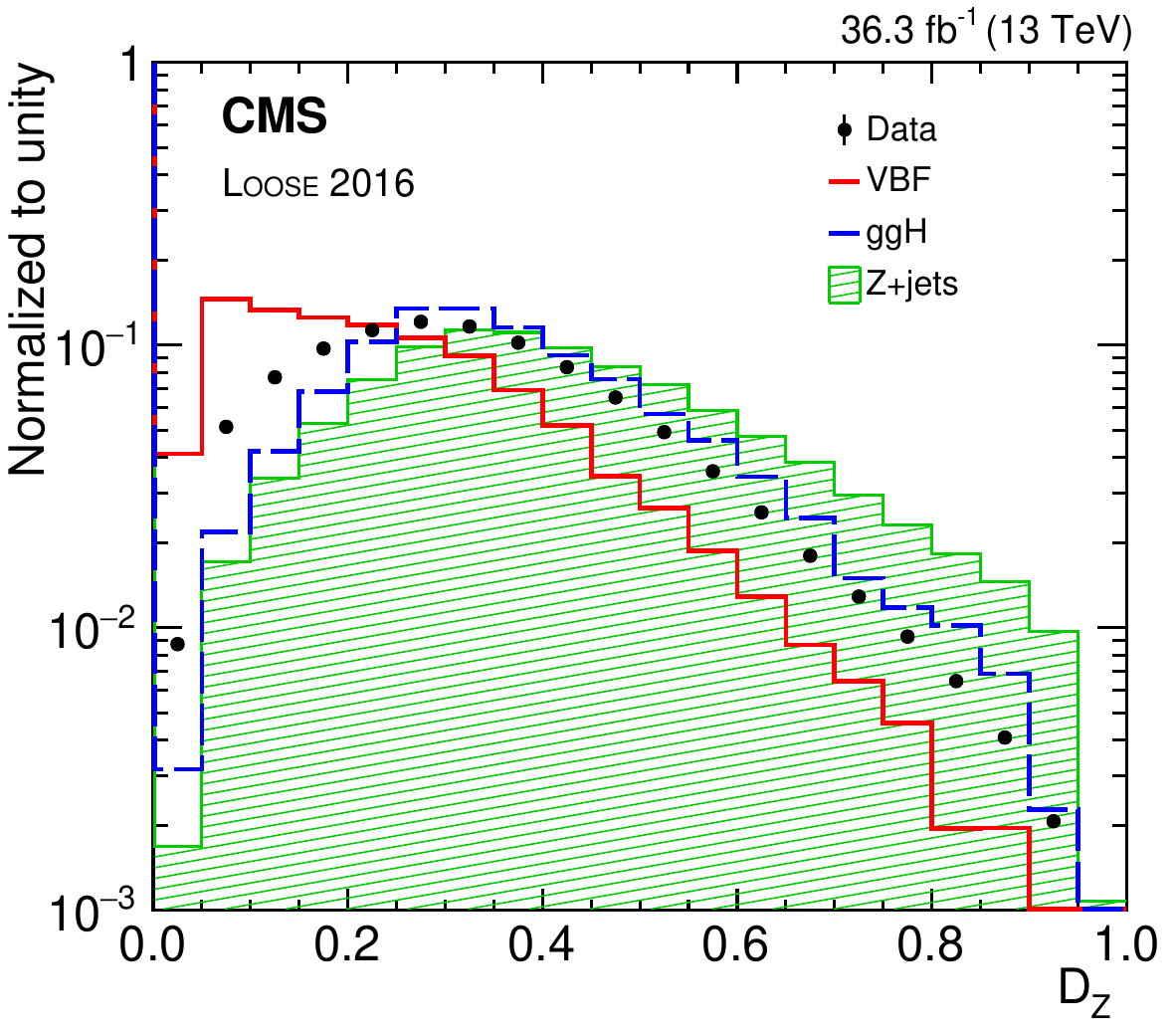}
        \includegraphics[width=.47\linewidth]{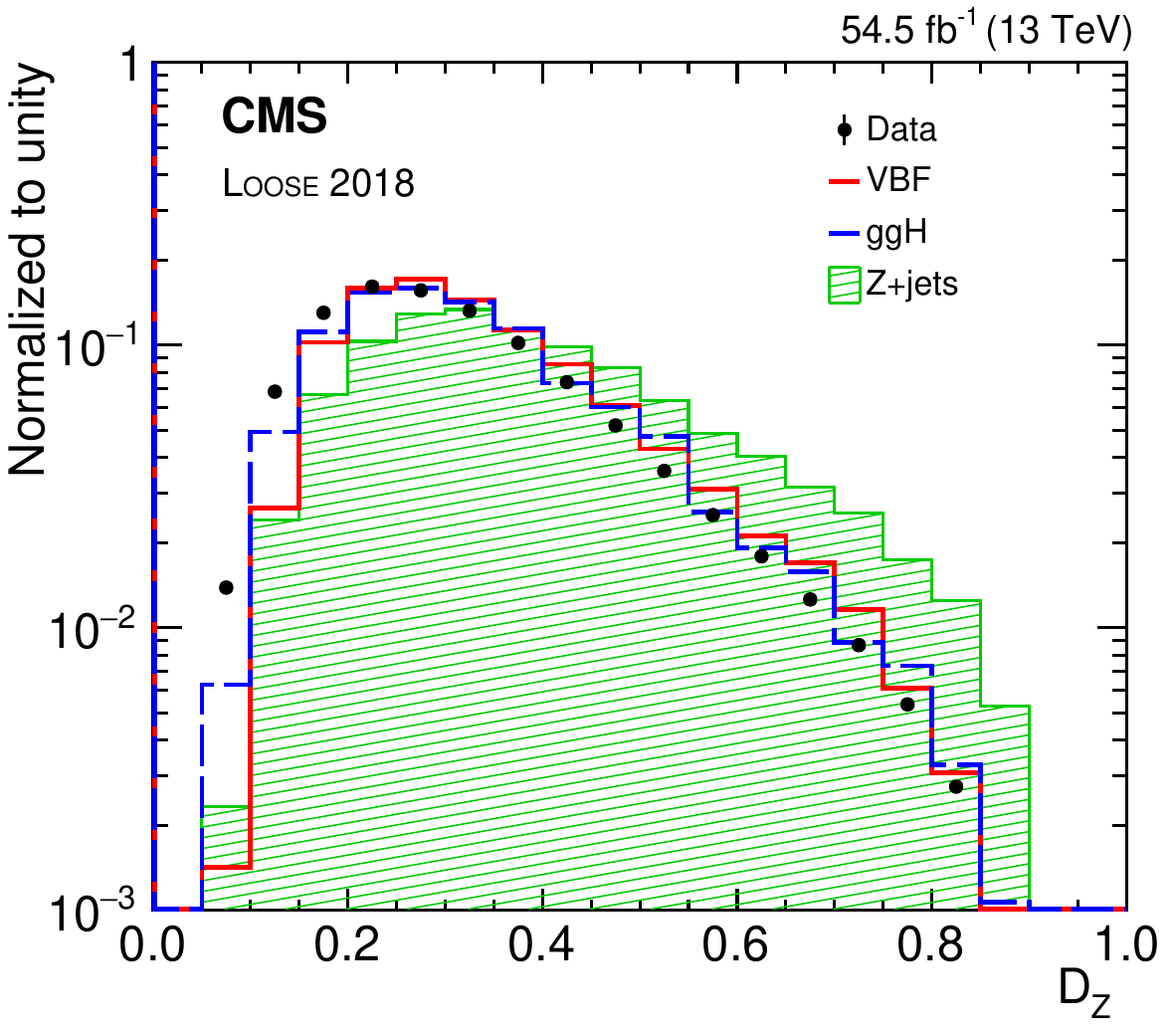}
        \caption{
                The unit normalized distributions of the BDT outputs: $D_{\ggF}$ (upper), $D_{\VBF}$ (middle), and $D_{\PZ}$ (lower) in data and simulated samples in the \LooseVBFA (left) and \LooseVBFB (right) analysis samples.
                Data events (points), dominated by the QCD multijet background, are compared to the VBF (red solid line), \ggF (blue dashed line), and \zjets (green hatched area) processes.
        }
        \label{fig:mva_loose}
\end{figure}

In the \LooseVBF samples, where larger contributions from \ggF and \zjets events are expected compared to the \TightVBF sample, a multiclass BDT is trained to separate four processes simultaneously: (1) VBF Higgs boson production, (2) \ggF Higgs boson production, (3) \zjets, and (4) QCD multijet events.
The training is performed using simulated samples of VBF, \ggF, and \zjets events and 5\% of the data in the \LooseVBFA and \LooseVBFB samples.
The BDTs trained in the \LooseVBF sample use the same set of input variables as the binary BDT classifiers trained in the \TightVBF analysis samples.
The multiclass BDT has four outputs, $D_{\VBF}$, $D_{\ggF}$, $D_{\PZ}$, and $D_{\QCD}$, quantifying the compatibility of each event with the VBF, \ggF, \zjets, and QCD multijet production hypotheses, respectively.
The distributions of $D_{\VBF}$, $D_{\ggF}$ and $D_{\PZ}$ normalized to unity are shown in Fig.~\ref{fig:mva_loose} for data and simulated samples.
Better discrimination between the VBF signal and background processes is observed in the \LooseVBFA sample compared to the \LooseVBFB sample, as visible in the $D_{\VBF}$ and $D_{\PZ}$ distributions, because of the different trigger requirements.
These samples contain events failing the \HLTTightVBF trigger.
In 2016 data, the trigger selection on \mjj and \detajj is more stringent compared with 2018 data (see Table~\ref{tab:Tab_evt_Sel}).
As a consequence, in the \LooseVBFB sample, a larger part of phase space, where \mjj and \detajj are particularly sensitive to the VBF signal, is removed in comparison with the \LooseVBFA sample, thus making the BDT in the \LooseVBFB sample less performant in discrimination between the VBF signal and background processes than in the \LooseVBFA sample.

Based on their BDT scores, events are classified into multiple exclusive categories, targeting the VBF, \ggF, and \zjets processes.
Details of the event categorization and the naming convention of the event categories are given in Table~\ref{tab:categories}.
The BDT thresholds defining each category, are optimized by maximizing the quantity $S/\sqrt{B}$, where $S$ is the number of expected events of the targeted process and $B$ is the number of QCD multijet background events approximated as the observed data events in the \mbb interval populated by the targeted process.
The \mbb interval of 80--100\GeV is used for the \zjets process and 104--146\GeV for the VBF and \ggF processes.
In total, 18 categories are introduced, three in each of the two \TightVBF analysis samples, and six in each of the two \LooseVBF analysis samples.

\begin{table}[htb!]
        \centering
        \topcaption{
                Event categorization used in the analysis for a total of 18 categories.
                The names of the categories are given in the first column.
                The BDT score boundaries defining each category are given in the second column and the targeted process is indicated in the third column.
        }
        \label{tab:categories}
        \begin{tabular}{lcc}
                Category        & BDT score boundaries                                    & Targeted process \\\hline
                \multicolumn{3}{c} {2016}                                                                    \\[\cmsTabSkip]
                \tabTightAOne   & $0.25< D<0.50$                                          & VBF              \\
                \tabTightATwo   & $0.50< D<0.75$                                          & VBF              \\
                \tabTightAThree & $0.75< D$                                               & VBF              \\[\cmsTabSkip]
                \tabLooseAOne   & $0.50< D_{\ggF}<0.55$                                   & \ggF             \\
                \tabLooseATwo   & $0.55< D_{\ggF}$                                        & \ggF             \\[\cmsTabSkip]
                \tabLooseAThree & $D_{\ggF}<0.50$, $0.80< D_{\VBF}<0.85$                  & VBF              \\
                \tabLooseAFour  & $D_{\ggF}<0.50$, $0.85< D_{\VBF}$                       & VBF              \\[\cmsTabSkip]
                \tabLooseAFive  & $D_{\ggF}<0.50$, $D_{\VBF}< 0.80$, $0.60< D_{\PZ}<0.75$ & \zjets           \\
                \tabLooseASix   & $D_{\ggF}<0.50$, $D_{\VBF}< 0.80$, $0.75< D_{\PZ}$      & \zjets           \\[\cmsTabSkip]
                \multicolumn{3}{c} {2018}                                                                    \\[\cmsTabSkip]
                \tabTightBOne   & $0.25< D<0.50$                                          & VBF              \\
                \tabTightBTwo   & $0.50< D<0.75$                                          & VBF              \\
                \tabTightBThree & $0.75< D$                                               & VBF              \\[\cmsTabSkip]
                \tabLooseBOne   & $0.55< D_{\ggF}<0.60$                                   & \ggF             \\
                \tabLooseBTwo   & $0.60< D_{\ggF}$                                        & \ggF             \\[\cmsTabSkip]
                \tabLooseBThree & $D_{\ggF}<0.55$, $0.50< D_{\VBF}<0.55$                  & VBF              \\
                \tabLooseBFour  & $D_{\ggF}<0.55$, $0.55< D_{\VBF}$                       & VBF              \\[\cmsTabSkip]
                \tabLooseBFive  & $D_{\ggF}<0.55$, $D_{\VBF}<0.50$, $0.60< D_{\PZ}<0.70$  & \zjets           \\
                \tabLooseBSix   & $D_{\ggF}<0.55$, $D_{\VBF}<0.50$, $0.70< D_{\PZ}$       & \zjets
        \end{tabular}
\end{table}

The aim of introducing distinct categories sensitive to the production of the \PZ boson is two-fold.
Firstly, these categories are intended to establish the signal from the \Zbb production, thereby validating the analysis techniques employed in this study.
Secondly, the tail of each \mbb distribution in the \zjets sample extends to the region partially populated by the signal events, thus affecting the precision of the measurement.
It is therefore important to constrain the background from the \zjets process with dedicated categories, thus improving the sensitivity to the signal.
Further, separate categories targeting the \ggF process improves the sensitivity of the analysis to inclusive Higgs boson production.

The numbers of selected events in data along with the expected background and signal yields in each analysis category are detailed in Tables~\ref{tab:yields1}~and~\ref{tab:yields2}.
The signal contribution is significantly smaller than the backgrounds in all analysis categories.

\begin{table}[htpb]
        \centering
        \topcaption{
                Event yields for various categories of the analyzed 2016 data corresponding to 36.3\fbinv, compared to the expected number of events from the simulated samples of signal and background other than the QCD multijet process.
                The quoted uncertainties are statistical only.
        }
        \label{tab:yields1}
        \begin{tabular}{ccccccc}
                Category        & VBF           & \ggF         & \zjets      & \ttbar \& $\PQt/\PAQt$ & \wjets       & Data    \\\hline
                \tabLooseAOne   & $4.5\pm0.2$   & $27.5\pm1.2$ & $275\pm16$  & $116.7\pm2.6$          & $10.6\pm3.0$ & $41432$ \\
                \tabLooseATwo   & $6.1\pm0.3$   & $51.6\pm1.8$ & $407\pm19$  & $127.3\pm2.7$          & $8.2\pm2.4$  & $58895$ \\
                \tabLooseAThree & $19.9\pm0.4$  & $2.7\pm0.4$  & $45\pm2$    & $10.6\pm0.8$           & $1.9\pm1.3$  & $4330$  \\
                \tabLooseAFour  & $17.4\pm0.4$  & $1.7\pm0.3$  & $31\pm5$    & $4.7\pm0.4$            & $0.5\pm0.3$  & $1901$  \\
                \tabLooseAFive  & $9.6\pm0.3$   & $34.8\pm1.4$ & $1150\pm20$ & $226.3\pm3.3$          & $40.8\pm6.1$ & $78850$ \\
                \tabLooseASix   & $3.1\pm0.2$   & $14.1\pm0.7$ & $650\pm10$  & $199.0\pm3.0$          & $35.0\pm5.9$ & $29992$ \\[\cmsTabSkip]
                \tabTightAOne   & $92.7\pm1.0$  & $15.6\pm1.0$ & $161\pm8$   & $37.6\pm1.4$           & $7.7\pm2.5$  & $29864$ \\
                \tabTightATwo   & $136.2\pm1.5$ & $13.9\pm0.9$ & $151\pm6$   & $22.5\pm1.2$           & $4.2\pm1.4$  & $21831$ \\
                \tabTightAThree & $117.3\pm1.1$ & $6.2\pm0.7$  & $75\pm3$    & $5.5\pm0.5$            & $3.1\pm1.1$  & $7231$
        \end{tabular}
\end{table}

\begin{table}[htpb]
        \centering
        \topcaption{
                Event yields for various categories of the analyzed 2018 data corresponding to 54.5\fbinv, compared to the expected number of events from the simulated samples of signal and background other than the QCD multijet process.
                The quoted uncertainties are statistical only.
        }
        \label{tab:yields2}
        \begin{tabular}{ccccccc}
                Category        & VBF           & \ggF         & \zjets     & \ttbar \& $\PQt/\PAQt$ & \wjets       & Data    \\\hline
                \tabLooseBOne   & $2.4\pm0.2$   & $13.5\pm2.1$ & $137\pm9$  & $32.4\pm1.4$           & $2.0\pm0.5$  & $17296$ \\
                \tabLooseBTwo   & $2.9\pm0.2$   & $24.3\pm1.5$ & $180\pm10$ & $33.1\pm1.4$           & $2.1\pm0.5$  & $24882$ \\
                \tabLooseBThree & $6.4\pm0.3$   & $1.0\pm0.3$  & $22\pm3$   & $4.4\pm0.5$            & $0.5\pm0.3$  & $1914$  \\
                \tabLooseBFour  & $11.0\pm0.4$  & $1.7\pm0.4$  & $25\pm3$   & $4.7\pm0.5$            & $0.3\pm0.2$  & $2453$  \\
                \tabLooseBFive  & $7.0\pm0.3$   & $10.8\pm0.9$ & $506\pm10$ & $59.0\pm1.7$           & $19.5\pm2.7$ & $24559$ \\
                \tabLooseBSix   & $4.0\pm0.2$   & $7.1\pm0.7$  & $445\pm7$  & $99.0\pm2.2$           & $24.0\pm2.8$ & $14530$ \\[\cmsTabSkip]
                \tabTightBOne   & $89.5\pm1.1$  & $18.0\pm3.2$ & $190\pm8$  & $49.3\pm1.5$           & $8.0\pm2.5$  & $29261$ \\
                \tabTightBTwo   & $139.8\pm1.4$ & $17.9\pm3.3$ & $202\pm7$  & $31.6\pm1.3$           & $7.2\pm2.4$  & $23392$ \\
                \tabTightBThree & $134.5\pm1.4$ & $8.7\pm2.6$  & $104\pm5$  & $7.6\pm0.6$            & $3.4\pm1.6$  & $8202$
        \end{tabular}
\end{table}

\subsection{Signal extraction}
\label{sec:extraction}

\begin{figure} [htb!]
        \centering
        \includegraphics[width=.45\linewidth]{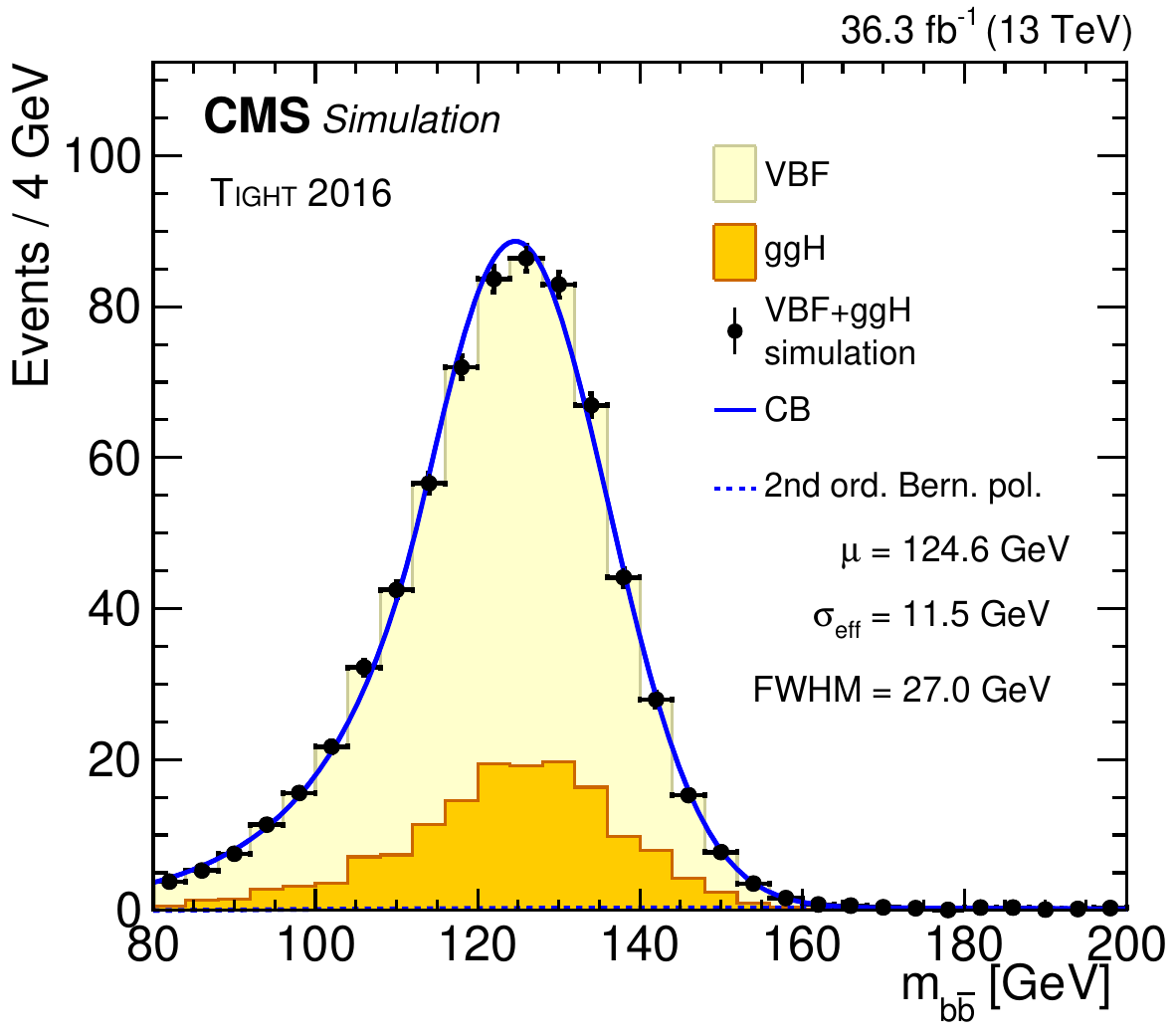}
        \includegraphics[width=.45\linewidth]{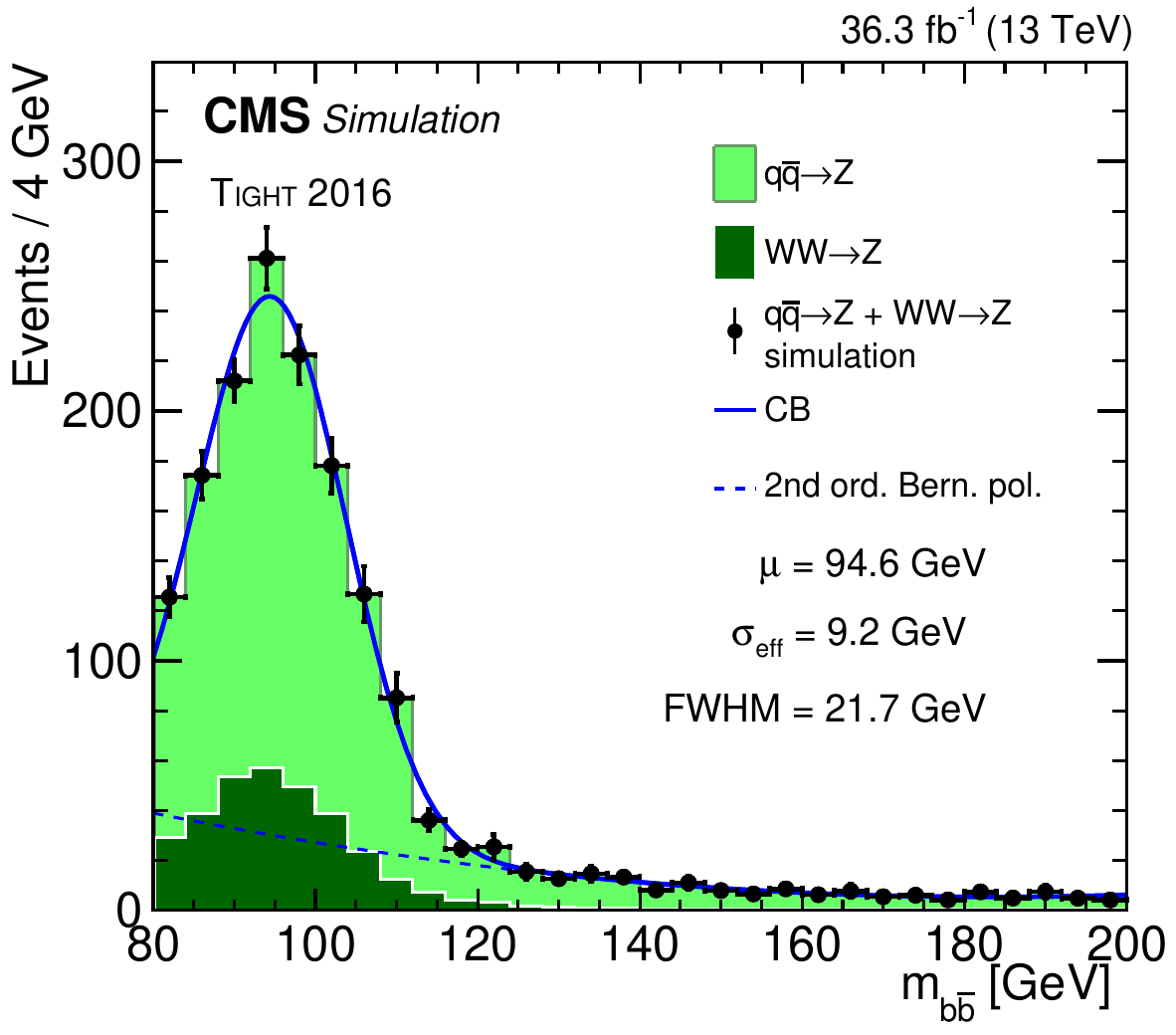}
        \caption{
                The \mbb distributions from simulation with overlaid parametric fits (solid blue lines) for the \TightVBFA analysis sample.
                Left: The fitted \mbb distribution in the signal combining the VBF (yellow histogram) and \ggF (orange) contributions.
                The black points refer to the total Higgs boson contribution from VBF and \ggF production modes.
                Right: The fitted \mbb distribution in simulated \zjets background (black points) combining the \WWZ (dark green histogram) and \qqZ (light green histogram) production modes.
                The black points refer to the total \zjets contribution from \qqZ and \WWZ modes.
                The dotted lines represent the second-order Bernstein polynomial components used to approximate the contributions from the wrong jet pairing.
        }
        \label{fig:sig_mod}
\end{figure}

The test statistic chosen to determine the signal yield is based on the profile likelihood ratio~\cite{LHCCLs}.
The signal is extracted from the simultaneous binned maximum likelihood fit of the \mbb distribution in all categories obtained from the data.
In each category, the \mbb distribution is fitted with a superposition of three parametric analytic functions accounting for: (i) the continuum background, dominated by QCD multijet events, (ii) resonant \zjets background and, (iii) signal.
Contributions from other processes are found to have a negligible effect on the results.
The addition of separate templates accounting for the \wjets, \ttbar, and $\PQt/\PAQt$ backgrounds, changes the extracted signal yield and its uncertainty by less than 1\%.
The aggregate contribution from other decays of the Higgs boson (\Hcc, $\PH\to\PGt\PGt$) and from the other production modes (\ttH and \VH) is evaluated to be almost two orders of magnitude lower than the contributions from the VBF and \ggF processes, hence their contributions are neglected in the signal extraction procedure.
The expected yields of the VBF, \ggF, and \zjets processes in a given category are estimated from simulation.
The \mbb distributions for the VBF and \ggF signal and \zjets background are individually modeled with a superposition of a one-sided Crystal Ball (CB) function~\cite{CrystalBallRef} for the resonant part and a second-order Bernstein polynomial accounting for the contribution from events where the jets are misassigned to the decay of the \PZ or the Higgs boson.
The parameters of the CB function and the Bernstein polynomial are extracted from the fit to the \mbb spectrum in the respective simulated sample, and separately for each sample.
For the signal, as well as the \zjets processes, the variations of the shapes across the different event categories within the same analysis sample are found to be well within the statistical uncertainties of the MC samples, and significantly smaller than the variations caused by the experimental uncertainties associated with the \PQb jet energy scale and resolution.
For the VBF and \ggF processes, which have similar shapes, a common analytic function is used to model the \mbb distribution in each analysis sample, namely \TightVBFA,  \LooseVBFA, \TightVBFB and \LooseVBFB.
The same approach is pursued for the \qqZ and \WWZ processes.
Figure~\ref{fig:sig_mod} illustrates the modeling of the \mbb shape in samples of the simulated VBF, \ggF, and \zjets events assigned to the \TightVBFA analysis sample.

For all analysis samples, the fitted peak positions of the CB functions describing the \mbb distribution in signal, are found to be consistent within fit uncertainties with the nominal value of the Higgs boson mass.
For the resonant \Zbb background, we observe a bias in the peak position of about 3.0--3.5\GeV above the nominal value of the \PZ boson mass.
The effect is caused by the relatively high \pt thresholds imposed on jets in the HLT and offline selections.
These thresholds, optimized primarily for the selection of \Hbb decays, result in an upward shift of the peak position in the \Zbb sample, where a larger fraction of low-\pt \PQb jets is expected compared to \Hbb decays.
A dedicated simulation study has demonstrated that no bias is present in the \Zbb sample if no trigger requirements are applied as well as if jet \pt thresholds are relaxed in the offline selection.

The shape and normalization of the continuum background are estimated from the fit to data.
The shape of the continuum background is modeled individually in each event category $i$ by a product of exponential and polynomial functions of order $n$,

\begin{equation}
        F^{\QCD}_{i} = \exp\left(-b_{i}\mbb\right)\left(1+\sum_{j=1}^n a_{ij} \mbb^{j}\right).
        \label{eq:qcd_mod}
\end{equation}

The family of functions given by Eq.~(\ref{eq:qcd_mod}) includes the case corresponding to the zeroth order polynomial.
In each category, the choice of the polynomial function is guided by the combined fit of two sideband regions, $80<\mbb<104\GeV$ and $146<\mbb<200\GeV$.
Sequential fits with increasing polynomial order are performed followed by a Fisher $F$-test~\cite{ref:ftest} to select the function with the fewest number of parameters necessary to fit the data.
The contribution from the \zjets background is accounted for by performing fits with the superposition of two functions,

\begin{itemize}
        \item tested analytic function from the family defined by Eq.~\ref{eq:qcd_mod} modeling continuum background dominated by the QCD multijet process;
        \item and a combination of the CB function and a second order Bernstein polynomial for the \zjets background as described earlier in this section.
\end{itemize}

First, the selection of functions for the modeling of the continuum background is performed in the categories targeting the \zjets process: \tabLooseAFive and \tabLooseASix in the 2016 and 2018 samples.
Fits in each of the four aforementioned categories are performed by varying the normalization of the \zjets process in an unconstrained manner.
An additional degree of freedom is introduced to account for the unconstrained normalization of the \zjets process.
A dedicated statistical test demonstrated that the measured normalizations of the \zjets process across the four categories are consistent.
The $p$-value quantifying the compatibility of these measurements is found to be 0.33.
Once the functional form of the continuum background modeling in each category targeting the \zjets process is selected based on the $F$-test result, a combined fit is performed in these four categories with a common, unconstrained \zjets normalization parameter.

The selection of functions from the family given in Eq.~\ref{eq:qcd_mod} in the categories targeting VBF and \ggF processes is done by performing fits with a Gaussian constraint imposed on the normalization of the \zjets component.
This constraint is obtained from the combined fit in the four categories targeting the \zjets process.

Ideally, the final result should not depend on the choice of the parametric fit function, but in reality, the choice of the function to model the background may bias the extraction of the signal.
Therefore, a dedicated study is performed for each category to assess a possible bias in the signal extraction.
For each of the selected functions, various alternative models of the continuum background are used to generate an ensemble of pseudo-data, including the injected signal and the contribution from \zjets process.
The resulting \mbb spectra obtained from the pseudo-data are fitted using the nominal model of the continuum background and the distribution of the extracted signal yield is compared to the injected yield.
The study is performed using the following alternative models:
\begin{enumerate}
        \item Functions from the same family, given by Eq.~(\ref{eq:qcd_mod}), but with polynomials one or two orders higher than the nominal function.
        \item Inverse polynomials, $I(\mbb)=1/(1+\sum_{j=1}^{n}a_{j}\mbb^{j})$, with $n=2$ or 3.
\end{enumerate}

The injected signal yield is varied between 50--200\% of the yield predicted by the SM to investigate a possible dependence of the bias on the signal strength.
The background model chosen yields a maximum potential bias that does not exceed 10\% of the statistical uncertainty on the fitted signal rate in each individual event category, and has negligible impact on the measurement.
The systematic uncertainty related to the choice of the function used to model the continuum background has a subdominant effect on the final results.
The functional forms employed in the modeling of the continuum background in each event category are reported in Table~\ref{tab:functions}.

\begin{table} [htb!]
        \centering
        \topcaption{
                The functional forms used to fit the continuum component of the background in various analysis categories.
                The notation ``exp'' stands for the exponential function, ``exp$\cdot$pol1 (pol2)" denotes the product of an exponential function and a first-order (second-order) polynomial.
        }
        \label{tab:functions}
        \begin{tabular}{ccc}
                Category & \multicolumn{2}{c}{Samples}                  \\\hline
                         & \LooseVBFA                  & \LooseVBFB     \\[\cmsTabSkip]
                G1       & exp$\cdot$pol2              & exp$\cdot$pol1 \\
                G2       & exp$\cdot$pol2              & exp$\cdot$pol1 \\
                V1       & exp                         & exp            \\
                V2       & exp                         & exp            \\
                Z1       & exp$\cdot$pol1              & exp$\cdot$pol1 \\
                Z2       & exp$\cdot$pol1              & exp$\cdot$pol1 \\[\cmsTabSkip]
                         & \TightVBFA                  & \TightVBFB     \\[\cmsTabSkip]
                1        & exp                         & exp            \\
                2        & exp                         & exp            \\
                3        & exp                         & exp
        \end{tabular}
\end{table}

\section{Systematic uncertainties}
\label{sec:syst}

Several systematic uncertainties affect the overall normalizations and shapes of the \mbb distributions for signal and background.
Systematic uncertainties are incorporated in the signal extraction procedure via nuisance parameters with Gaussian or log-normal probability density functions (pdfs), which are treated according to the frequentist paradigm.
These pdfs are included as additional factors in the likelihood function, which is maximized in the fit.
All uncertainties are divided into two categories.
Theoretical uncertainties arise from the limited precision in the computation of the inclusive and differential cross sections of the modeled processes and are fully correlated between event categories and data-taking periods.
Experimental uncertainties comprise (i) uncertainties related to the imperfect simulation of the detector response and consequent inaccurate modeling of the reconstruction of physics objects and observables in the simulated samples, (ii) the uncertainty in the integrated luminosity estimation, and (iii) the uncertainty in the modeling of the pileup interactions.
The dominant uncertainties affecting this measurement are briefly summarized in the following.

\textbf{Theoretical uncertainties}

\begin{itemize}
        \item \textit{Parton shower uncertainty in \PYTHIA}: The uncertainty related to the modeling of the parton shower in \PYTHIA is estimated by varying the scales of the initial- and final-state QCD radiation.
              This uncertainty has a sizable effect on modeling of jet-related variables used as inputs to the BDT, thereby affecting acceptance of signal and background events in different event categories.
              The impact on the predicted event yields varies between 2 and 10\% across event categories with larger effects observed for categories targeting \VBF production.
              The modeling of key BDT input variables, such as \mjj and \detajj, is validated with a control sample of events, in which a \PZ boson is produced in association with at least two jets and decays into a pair of muons.
              Distributions of these variables in data and simulated samples agree within instrumental and theoretical uncertainties, including the uncertainty related to the modeling of the parton shower in \PYTHIA.
              Variations in the \mbb distribution within these uncertainties are found to be negligible in the simulated signal and \zjets background samples.
              Inclusion of these shape-altering effects in the statistical inference procedure changes the uncertainty in the extracted rates of the signal and resonant \Zbb background by less than 0.5\%.
        \item \textit{Parton shower and hadronization model for VBF production}: Simulation of the color connection between the incoming and outgoing partons in the VBF process is particularly sensitive to the choice of the event generator.
              The related uncertainty is assessed by comparing the dipole parton shower model implemented in \PYTHIA with the alternative one provided by the \HERWIG event generator.
              Yields of the \VBF signal in each event category are compared between the two models and the differences are used to define a double-sided uncertainty with the \PYTHIA prediction taken as the central estimate.
              The choice of the event generator only marginally affects the shapes of the \mbb distributions in different event categories.
              The corresponding shape variations in the \mbb spectra modify the expected uncertainty in the extracted rate of the \VBF process by less than 1\%.
              The choice of the generator also impacts the simulation of forward jets in the \WWZ sample.
              The validation study performed using the control sample of \PZ boson decays into muons, described above, demonstrates that properties of spectator jets accompanying the production of a \PZ boson, are well modeled by \PYTHIA.
              Small discrepancies between data and simulation observed in distributions of key variables are fully covered by variations of the scales of the initial- and final-state QCD radiation.
              Therefore, no additional uncertainty is assigned to the parton shower modeling in the simulated \WWZ sample.
        \item \textit{Scale variations}: Uncertainties arising from missing higher-order QCD corrections in the theoretical computations affect the overall production rates as well as the kinematics of the simulated processes.
              The impact of these uncertainties on the cross sections of the VBF and the \ggF productions are estimated by varying the renormalization and factorization scales, yielding uncertainties of $\pm$2.1\% and $^{+4.6}_{-6.7}$\%, respectively~\cite{LHCHiggsCrossSectionWorkingGroup:2016ypw}.
              The impact on the acceptance of VBF and \ggF events across the event categories is evaluated by applying event weights that reflect the effect of the scale variations on the kinematics of the simulated events.
              The variations in event yields across categories are typically between 1 and 4\%, with larger effects observed in categories targeting the VBF process.
        \item \textit{Uncertainties in PDFs and \alpS}: The PDF and \alpS uncertainties in the cross sections of the VBF and \ggF processes are $\pm$0.4\% and $\pm$3.2\%, respectively~\cite{LHCHiggsCrossSectionWorkingGroup:2016ypw}.
        \item \textit{Uncertainty in the branching fraction}: The theoretical uncertainty in the branching fraction of the Higgs boson decay to bottom quarks is 0.65\% at $m_{\PH}=125\GeV$~\cite{LHCHiggsCrossSectionWorkingGroup:2016ypw}.
        \item \textit{Uncertainties from \zjets NLO $K$-factors}: The MC samples of \qqZ and \WWZ events are simulated with \MGvATNLO at the LO accuracy with MLM matching.
              To improve the modeling of these processes, additional \pt-dependent $K$-factors are applied following the prescription in Ref.~\cite{Lindert:2017olm} to match the differential distribution of the \PZ boson \pt as predicted by the NNLO QCD and NLO EW calculations.
              The $K$-factors related to the NLO EW corrections are applied to both \qqZ and \WWZ samples.
              The $K$-factors associated with the NNLO QCD corrections are applied only to the \qqZ sample.
              The uncertainties in these $K$-factors are incorporated via six nuisance parameters.
              Three of these nuisance parameters correspond to the NLO EW corrections and affect both \qqZ and \WWZ samples, whereas another three parameters are related to the NNLO QCD corrections and have an impact only on simulated \qqZ events.
              Variations of these nuisance parameters change the acceptance of \zjets events in different event categories by 1--7\%.
              Uncertainties related to missing higher order QCD corrections in the simulated \WWZ sample are evaluated by varying renormalization and factorization QCD scales, modifying the acceptance of \zjets events by 1--3\% across different event categories.
\end{itemize}

\textbf{Experimental uncertainties}

\begin{itemize}
        \item \textit{Jet energy scale and resolution corrections}: The uncertainties in the jet energy scale and resolution originate from different sources with limited correlations~\cite{JetsInRun2,JECperformance}.
              The uncertainties depend on the jet kinematics and are typically larger in the forward regions.
              Variations of the jet four-momenta caused by these uncertainties have an impact on the acceptance in various event categories of between 5 and 10\%.
        \item \textit{\PQb tagging}: The uncertainties in \PQb tagging efficiency are evaluated with the control samples of the semileptonic \ttbar decays, \zjets events, and the inclusive QCD multijet events~\cite{BTV-16-002}.
              The uncertainties associated with the selection of the working point of the \DEEPJET score for the tagger vary between 4--8\% depending on jet flavor, \pt, and $\eta$.
              The \DEEPJET discriminants of the two \PQb-tagged jets are also used as an input to the BDT classifiers.
              An assessment of control samples of QCD multijet and leptonic \ttbar events revealed only small differences in the \DEEPJET discriminant distribution between data and simulated events above the threshold defining the working point.
              These differences are found to have subdominant effects on the response of BDT classifiers, leading to a migration of events between categories at a percent level.
        \item \textit{Trigger efficiency}: The corrections to the trigger efficiency are discussed in Section~\ref{sec:triggers}.
              The uncertainties in the trigger scale factors are predominantly statistical.
              The impact of these uncertainties on the acceptance of signal and background events in various analysis categories is estimated to range from 5 to 10\%.
        \item \textit{Corrections to \PQb jet energy regression}: The scale and smearing corrections to the regressed \PQb jet energy are derived as described in Section~\ref{sec:bregression}.
              The uncertainties in the corrections are propagated to the parameters of the CB functions modeling the core of the \mbb spectrum in the simulated samples of \Hbb and \Zbb decays.
              Furthermore, these uncertainties change the shape of several observables used as inputs to the BDT classifiers, and consequently affect the estimate of the acceptance in various event categories.
        \item \textit{Integrated luminosity}: The uncertainty in the luminosity measurement has both correlated and uncorrelated components, which vary across different data-taking periods~\cite{CMSlumi2016,CMSlumi2018}.
              The uncertainties of the individual years are 1.2 and 2.5\% for 2016 and 2018, respectively.
              The uncertainty of the combined 2016 and 2018 integrated luminosity is 1.7\% (calculated using the full correlation scheme and per-year luminosities of 36.3\fbinv and 54.5\fbinv, respectively).
        \item \textit{Pileup modeling}: The number of primary interactions per bunch crossing varies with the instantaneous luminosity during the data-taking operation.
              In order to match the distribution of pileup in data with that in the simulated samples, weights are applied that are determined by studying minimum-bias data sets.
              A normalization uncertainty is derived by altering the pileup weights obtained by changing the total inelastic cross section by $\pm$4.6\%~\cite{CMS:2018mlc} of its nominal value.
        \item \textit{Pileup jet identification}: This uncertainty is estimated by comparing the pileup jet identification score~\cite{CMS:2020ebo} in events with a \PZ boson decaying into a pair of electrons or muons and one balanced jet in data and simulation.
              The assigned uncertainty depends on \pt and $\abs{\eta}$, and is designed to cover all differences between data and simulation in the distribution.
\end{itemize}

The impact of the most significant uncertainties is presented in Table~\ref{tab:systematics}.

\begin{table} [htb!]
        \centering
        \topcaption{
                The impact of the dominant systematic uncertainties on the observed signal strength for inclusive Higgs boson production followed by decay to bottom quarks.
        }
        \label{tab:systematics}
        \centering
        \begin{tabular}{cc }
                Source of systematic uncertainty      & Impact on signal strength [\%] \\
                \hline
                VBF parton shower                     & 13.0                           \\
                Jet energy scale                      & 7.7                            \\
                Trigger efficiency                    & 6.7                            \\
                Parton shower (final-state radiation) & 5.6                            \\
                \PQb jet regression smearing          & 3.3                            \\
                \PQb tagging efficiency               & 3.0                            \\
                Pileup modeling                       & 2.3                            \\
                \PQb jet regression scale             & 2.0                            \\
                Jet energy resolution                 & 1.5                            \\
        \end{tabular}
\end{table}

\section{Results}
\label{sec:results}

This analysis is primarily sensitive to the VBF Higgs boson production followed by \Hbb decay.
The outcome of the measurement depends on how the contribution from the \ggF process is accounted for, and three scenarios are presented that differ in the way this process is treated in the signal extraction procedure.
In each scenario, the best fit value of the parameters of interest and their confidence level (CL) intervals are extracted following the procedure described in Section~3.2 of Ref.~\cite{Khachatryan:2014jba}.

\subsection{Measurement of inclusive Higgs boson production}

In the measurement of the inclusive Higgs boson production rate, the \ggF process is considered as part of the signal.
The fit is performed with an unconstrained signal strength modifier \muHbb that simultaneously scales the yields of VBF and \ggF events in all categories.
The parameter \muHbb is the product of the inclusive production cross section of the Higgs boson and the \Hbb branching fraction relative to the SM expectation.
In the measurement, we allow the overall normalization of the \zjets process \muZ to vary in an unconstrained manner.

The analytic function used to fit the \mbb spectrum in the $i$th category is given by
\begin{equation}
\begin{aligned}
        F_{i}(\mbb|\vec{\theta}) & =  N_{i}^{\QCD}F_{i}^{\QCD}(\mbb|\vec{\alpha}_i) + \muZ N_{i}^{\zbb}(\vec{\theta}) F_{i}^{\zbb}(\mbb|\vec{\theta}_S)\\
                                 & + \muHbb (N_{i}^{\qqH}(\vec{\theta}) + N_{i}^{\ggF}(\vec{\theta})) F_{i}^{\hbb}(\mbb|\vec{\theta}_S).
\end{aligned}
\end{equation}

The function includes the following category-dependent components:

\begin{itemize}
        \item $N_{i}^{\QCD}$: the normalization of the QCD multijet background extracted from the fit.
        \item $F_{i}^{\QCD}(\mbb|\vec{\alpha}_i)$: analytic parametric function normalized to unity modeling the shape of the QCD multijet background.
              The parameters $\vec{\alpha}_i$ of the function are obtained from the fit.
              The types of functions used to model the QCD multijet background are discussed in Section~\ref{sec:extraction}.
        \item $N_{i}^{\qqH,\ggF,\zbb}(\vec{\theta})$: predicted yields of the VBF, \ggF, and \zjets processes; these yields depend on the nuisance parameters $\vec{\theta}$ that incorporate systematic uncertainties in the fit.
        \item $F_{i}^{\hbb,\zbb}(\mbb|\vec{\theta}_S)$: analytic functions normalized to unity modeling the shape of the \mbb distribution in the samples of \Hbb and \Zbb decays.
              The parameters of the analytic functions, described in Section~\ref{sec:extraction}, are influenced by the nuisance parameters $\vec{\theta}_S$ associated with the uncertainties in the scale and resolution of the \PQb jet energy regression.
\end{itemize}

The measurement of the inclusive Higgs boson production yields the best fit values as
\begin{align*}
        \muHbb & = 0.99^{+0.32}_{-0.20}\syst\pm 0.36\stat,\quad\text{and} \\
        \muZ   & = 0.96\pm0.21\syst\pm 0.23\stat.
\end{align*}

The \Hbb signal, including contributions from the VBF and \ggF processes, is observed with a statistical significance of 2.6\,$\sigma$, compared to the expected significance of 2.9\,$\sigma$.
The sensitivity of the measurement is driven by the \TightVBF categories, while the \LooseVBF categories constrain the \zjets background.
The expected uncertainty in $\muHbb$ is determined from the fit to the Asimov dataset~\cite{Cowan:2010js}, where cross sections of the \PZ and Higgs boson production are set to the values predicted by the SM, whereas parameters of pdfs describing the continuum background, are set to the values obtained from fits to sideband regions.
The expected systematic and statistical uncertainties are both found to be about 10\% smaller than the respective observed ones.
As a consequence, the expected significance is about 10\% higher than the observed one, despite the central observed value of $\muHbb$ being close to unity.
The linear correlation coefficient between $\muHbb$ and $\muZ$ is found to be $0.37$.
In the fit, yields of \Hbb and \Zbb events are both anti-correlated with the yield of a large continuum background.
Hence, a positive correlation between $\muHbb$ and $\muZ$ is expected and confirmed by the fit result.

\begin{figure} [htb!]
        \centering
        \includegraphics[width=.32\linewidth]{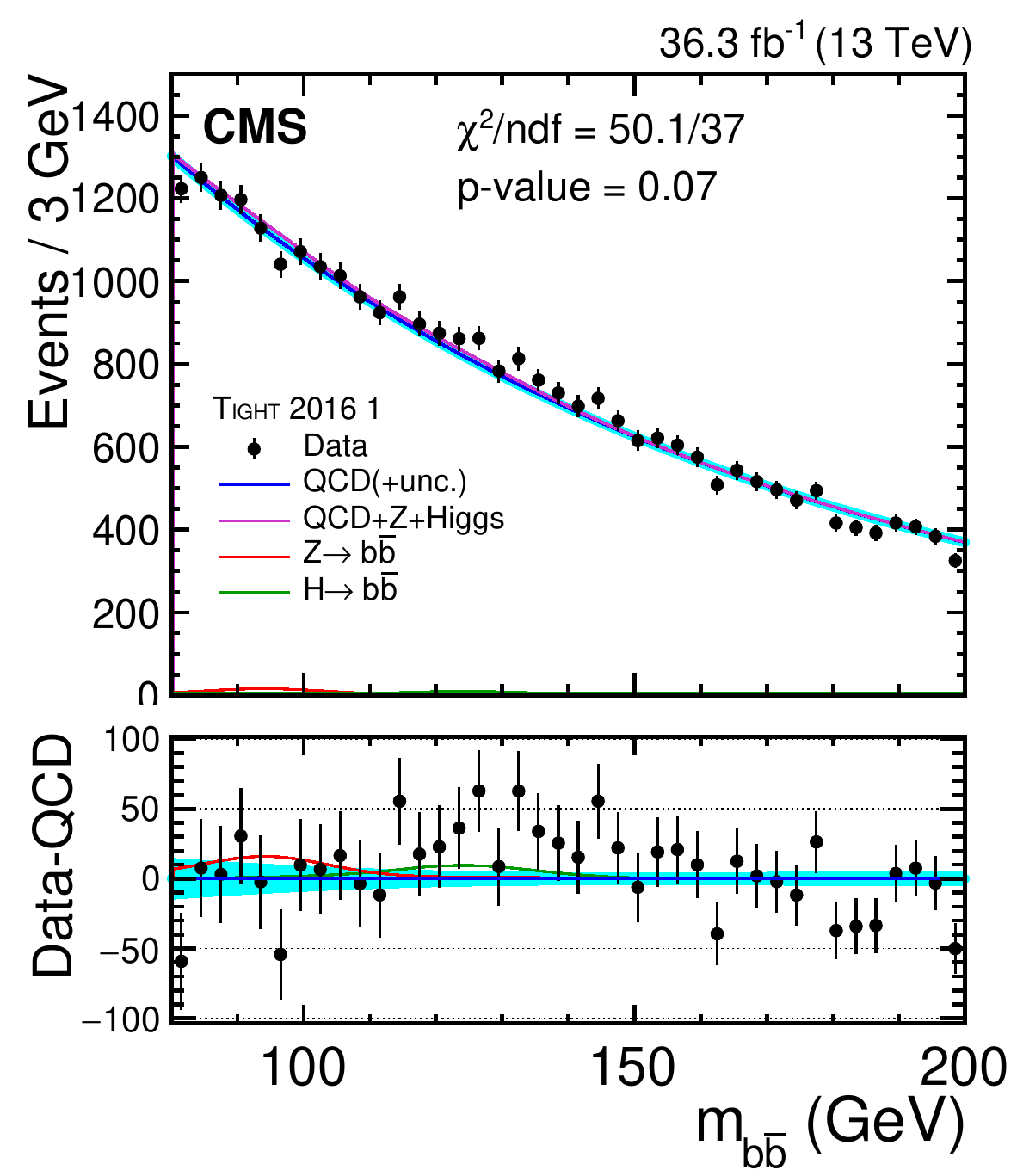}
        \includegraphics[width=.32\linewidth]{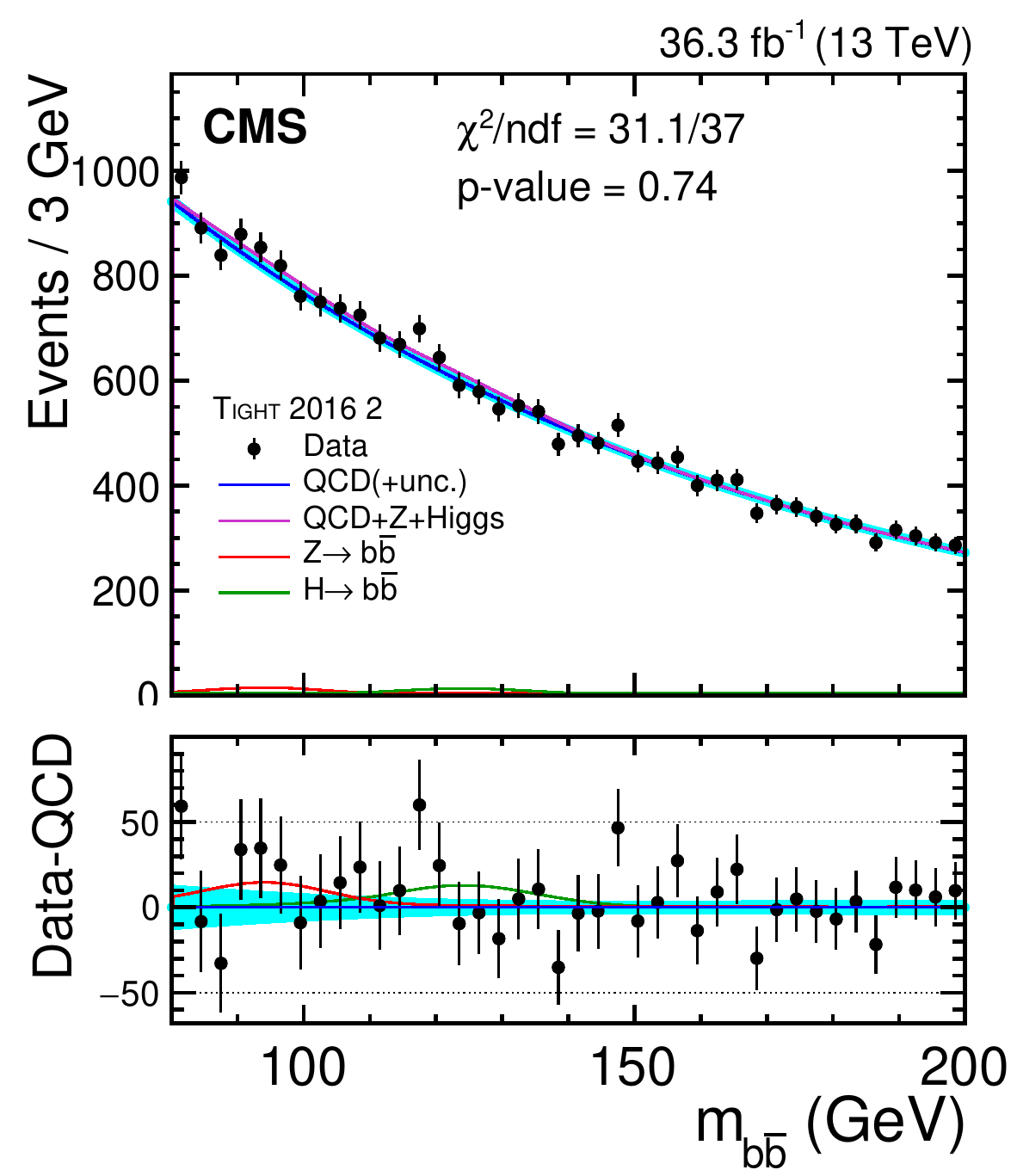}
        \includegraphics[width=.32\linewidth]{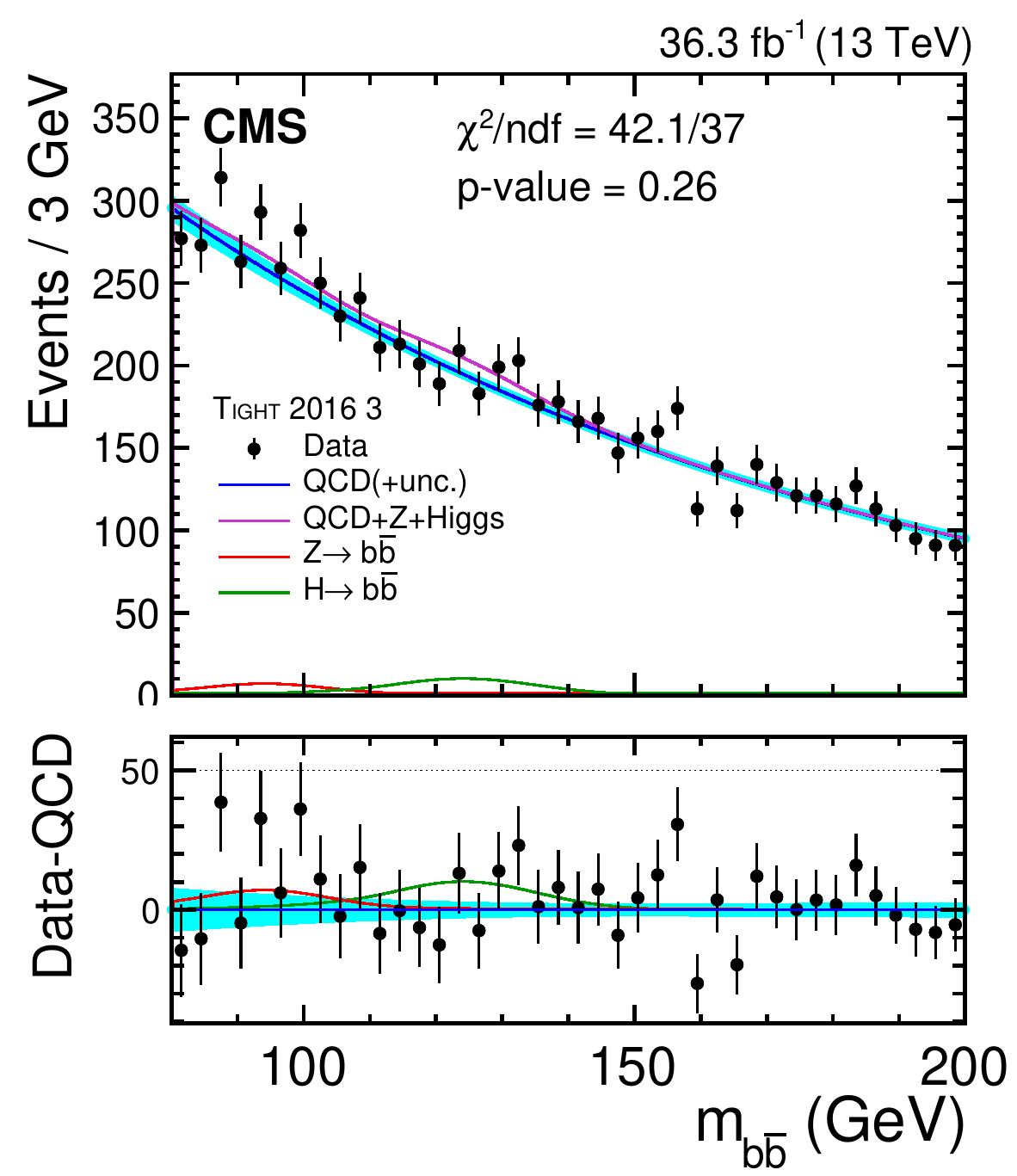}
        \caption{
                The \mbb distributions in three event categories: \TightAOne (left), \TightATwo (center), and \TightAThree (right).
                The black points indicate data, the blue solid curve corresponds to the fitted nonresonant component of the background, dominated by QCD multijet events, and the shaded (cyan) band represents the $\pm$1\,$\sigma$ uncertainty band.
                The total signal-plus-background model includes contributions from \Zbb, \Hbb, and the nonresonant component; it is represented by the magenta curve.
                The lower panel compares the distribution of the data after subtracting the nonresonant component with the resonant contributions of the \Zbb background (red curve) and \Hbb signal (green curve).
        }
        \label{fig:fit_sb_A}
\end{figure}

\begin{figure} [htb!]
        \centering
        \includegraphics[width=.32\linewidth]{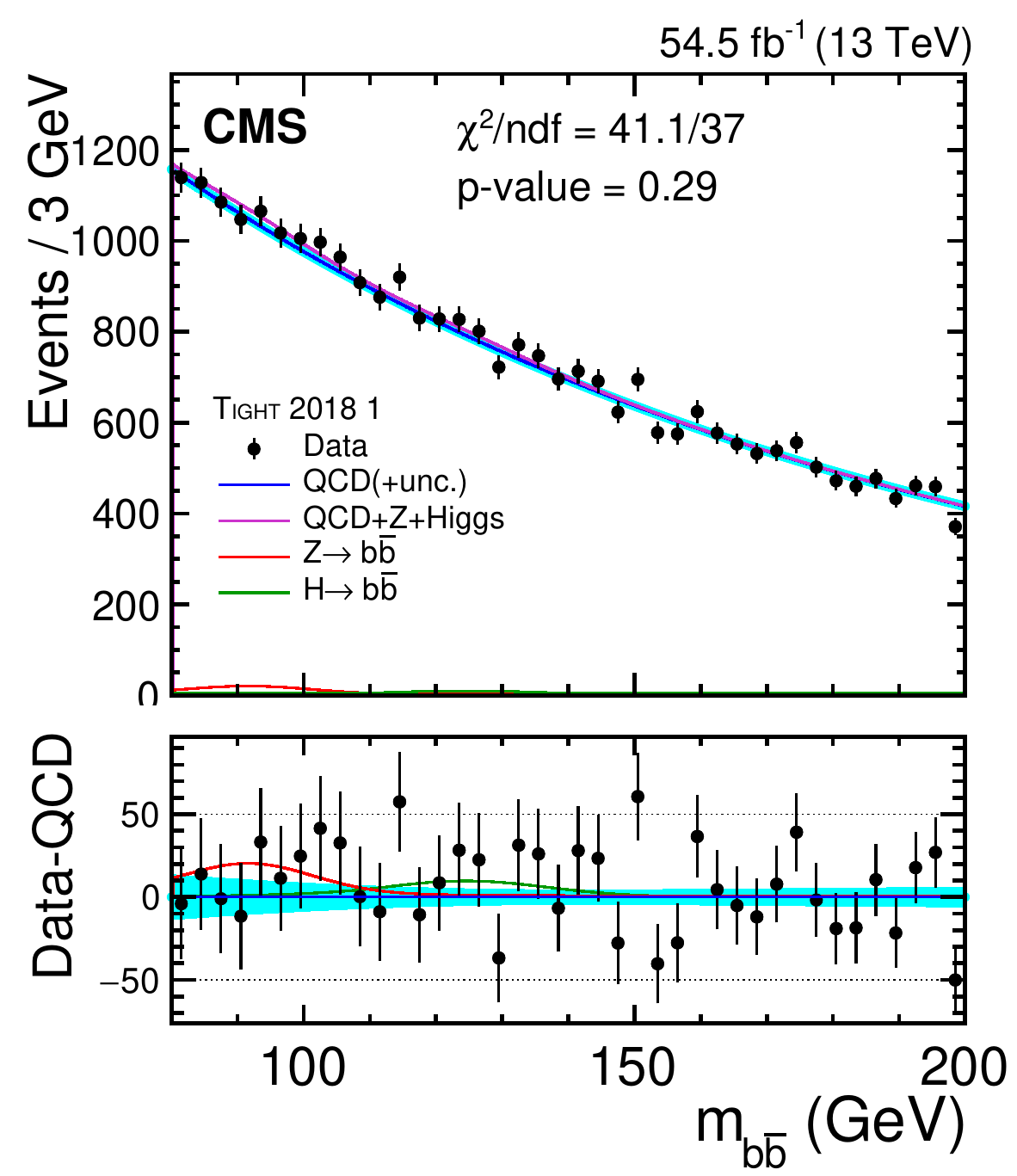}
        \includegraphics[width=.32\linewidth]{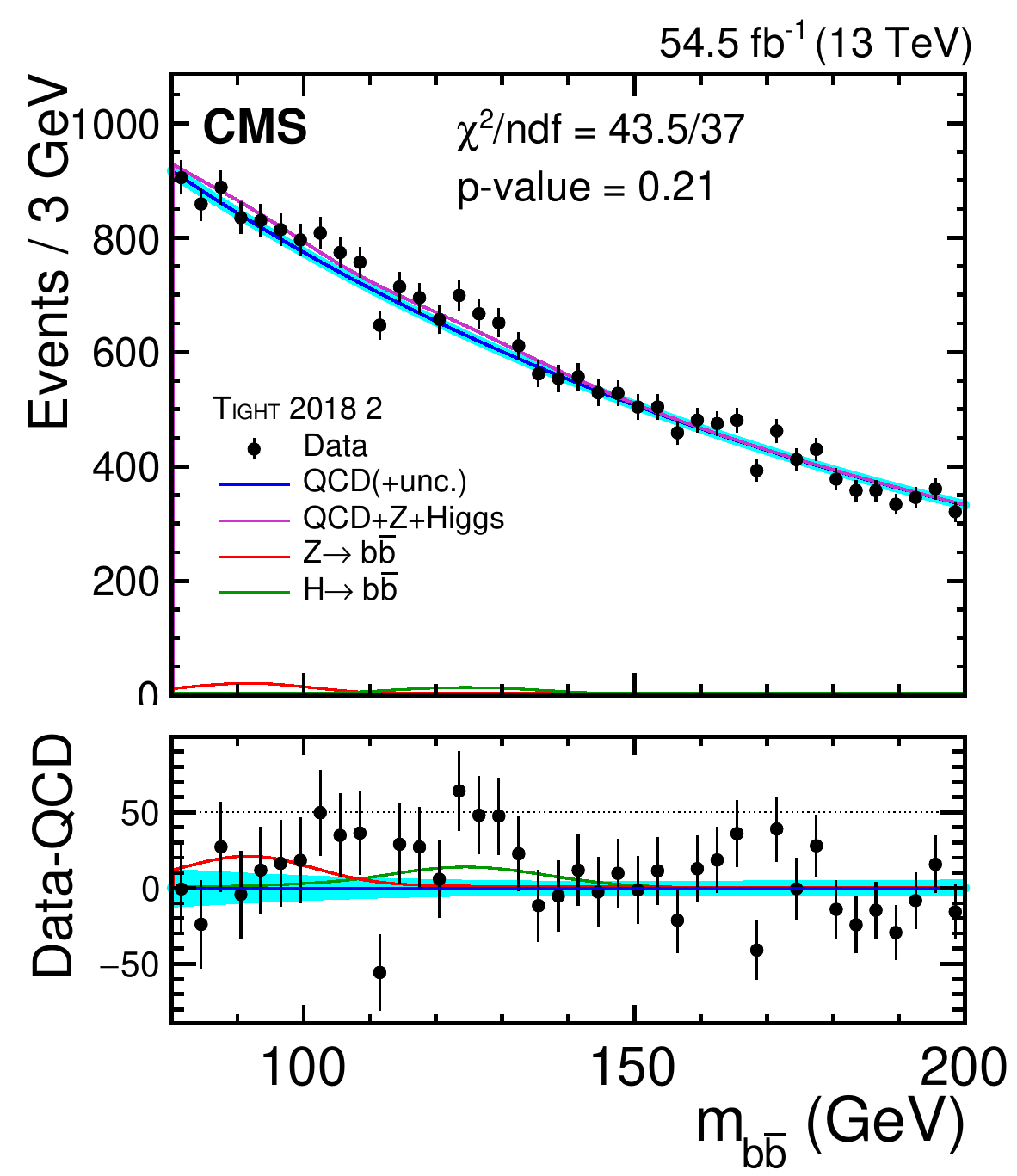}
        \includegraphics[width=.32\linewidth]{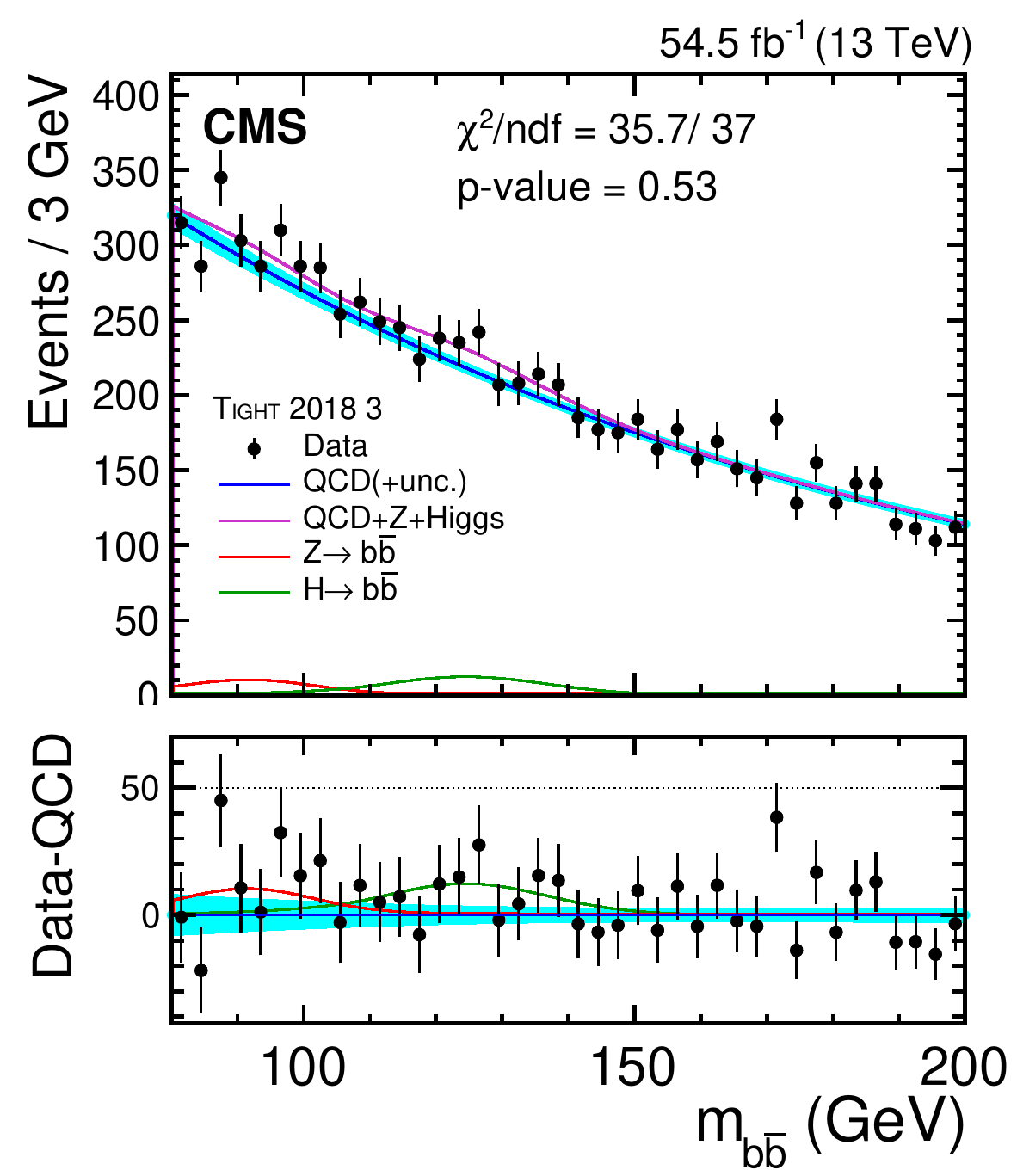}
        \caption{
                The \mbb distributions in three event categories: \TightBOne (left), \TightBTwo (center), and \TightBThree (right).
                A complete description is given in Fig.~\ref{fig:fit_sb_A}.
        }
        \label{fig:fit_sb_B}
\end{figure}

\begin{figure} [htb!]
        \centering
        \includegraphics[width=.40\linewidth]{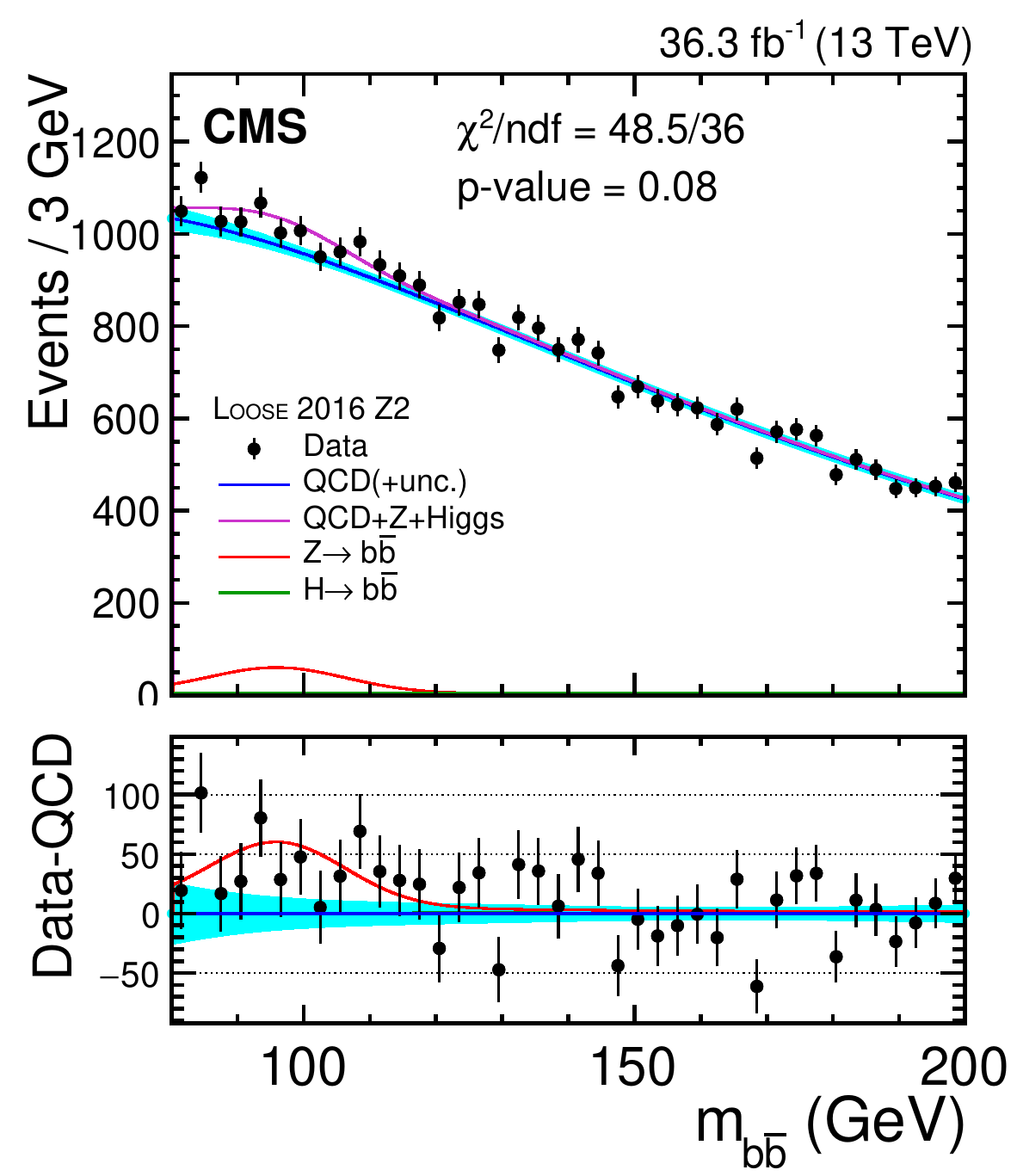}
        \includegraphics[width=.40\linewidth]{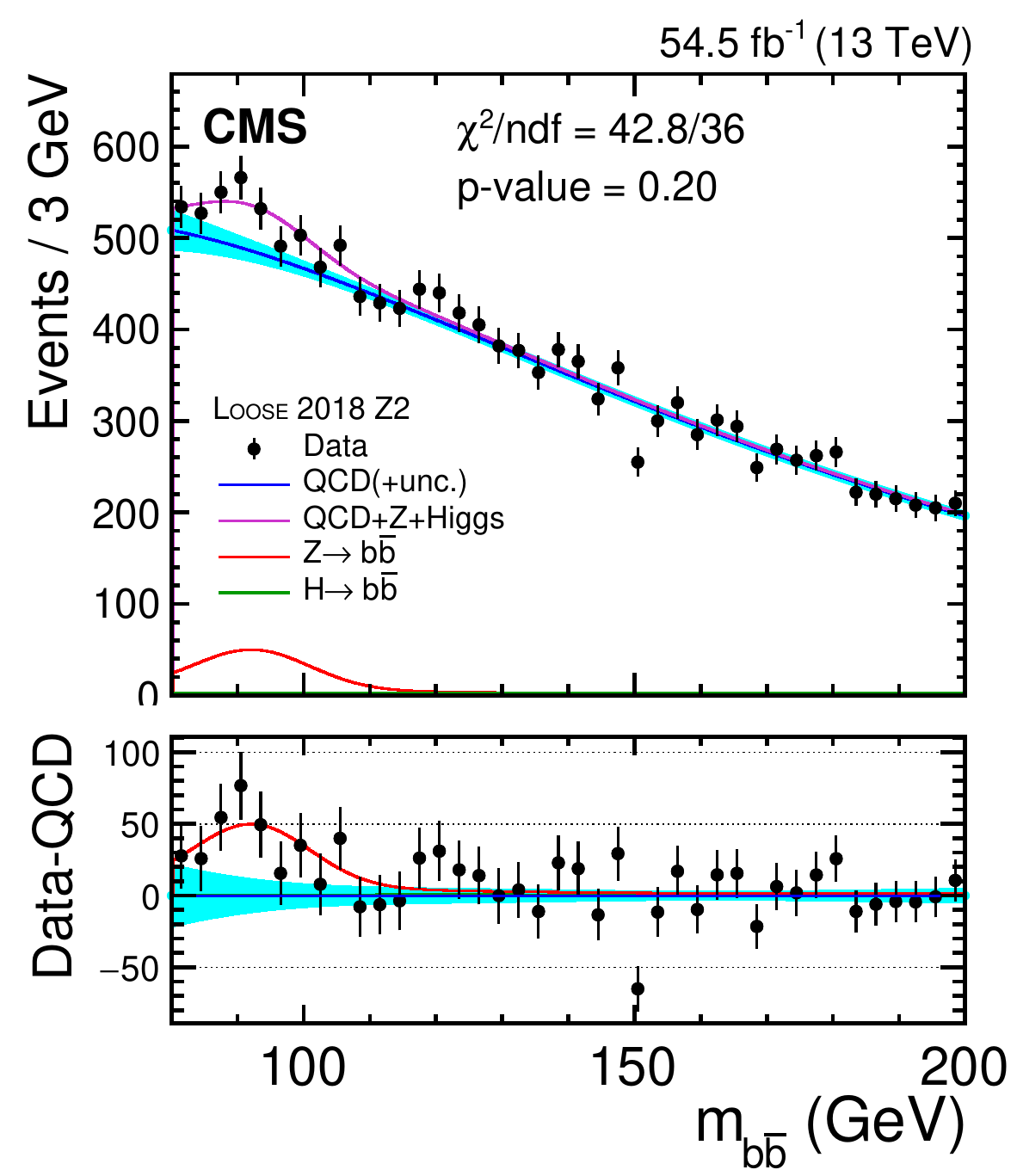}
        \caption{
                The \mbb distributions in two event categories: \LooseASix (left) and \LooseBSix (right).
                A complete description is given in Fig.~\ref{fig:fit_sb_A}.
        }
        \label{fig:fit_sb_ZJ}
\end{figure}

Figures~\ref{fig:fit_sb_A}~and~\ref{fig:fit_sb_B} show the results of the fit in the \TightVBFA and \TightVBFB categories, and Fig.~\ref{fig:fit_sb_ZJ} is for the \LooseASix and \LooseBSix categories.
In these Figures, $\chi^{2}$ quantifies the consistency of the \mbb distribution observed in data with the fitted parametric analytic function, ndf is the number of degrees of freedom in the fit, and the $p$-value is defined as the probability of observing the $\chi^{2}$ value larger than the actual one.

\begin{figure} [htb!]
        \centering
        \includegraphics[width=.5\linewidth]{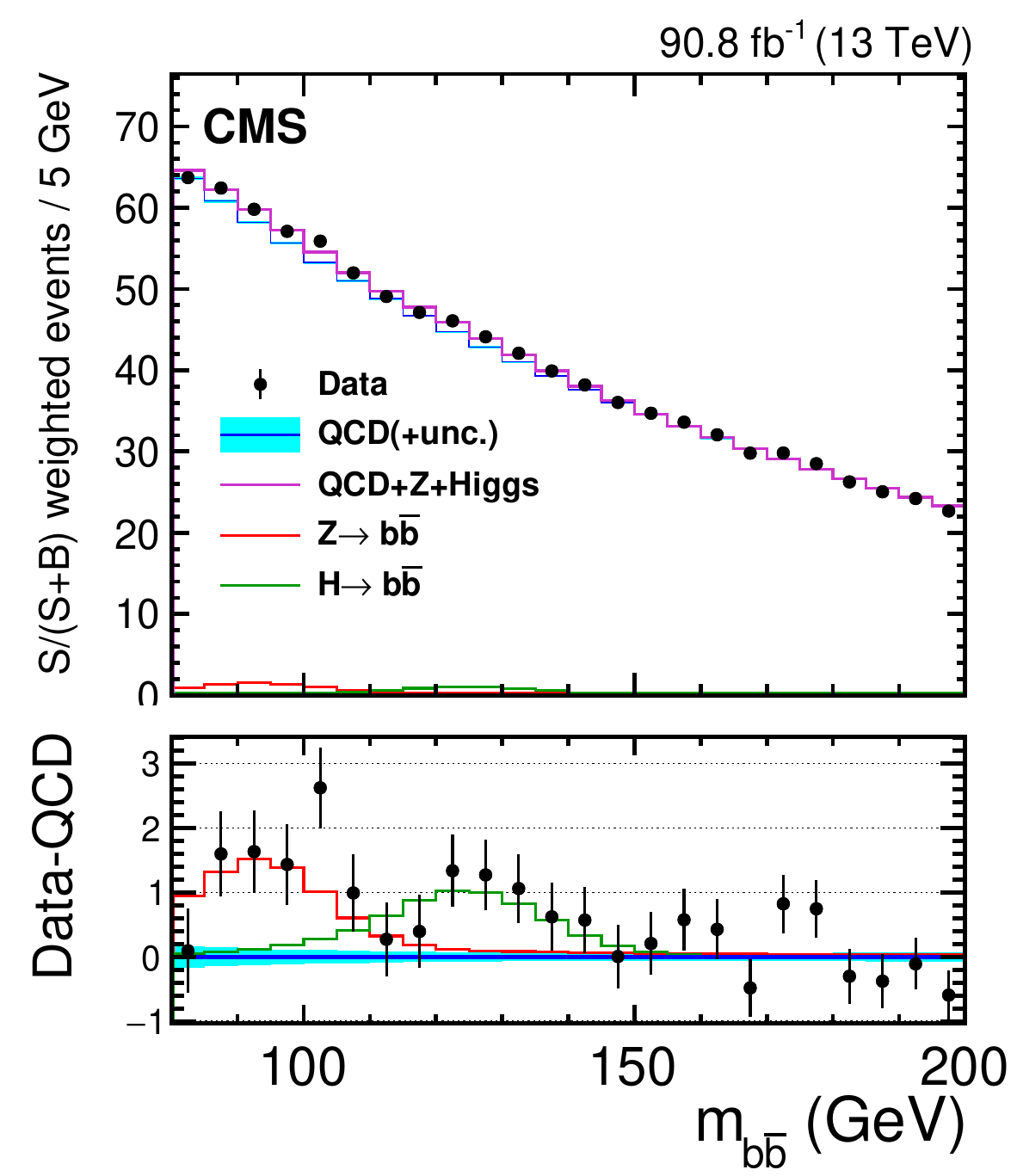}
        \caption{
                The \mbb distribution after weighted combination of all categories in the analysis weighted with $S/(S+B)$.
                A complete description is given in Fig.~\ref{fig:fit_sb_A}.
        }
        \label{fig:fit_sb_wt}
\end{figure}

Figure~\ref{fig:fit_sb_wt} shows the \mbb spectrum combining all 18 analysis categories.
Each category enters the combination with a weight $S/(S+B)$, where $S$ is the total \Hbb signal yield (VBF and \ggF contributions) and $B$ is the total background yield including QCD multijet and \zjets processes obtained in a given category from the combined fit.
The yields are computed by integrating the \mbb distribution over the entire fitted range of 80--200\GeV.
The distribution in data is compared with the fitted background-only model and the signal-plus-background model.

\subsection{Measurement of VBF production when \texorpdfstring{\ggF}{ggF} production is constrained to SM expectations}

The measurement of the exclusive VBF production rate is performed with the contribution from the \ggF process constrained within theoretical and experimental uncertainties to the SM expectation.
In this case, the analytic function employed to fit the \mbb spectrum in category $i$ is modified to be
\begin{equation}
\begin{aligned}
        F_{i}(\mbb|\vec{\theta}) & =  N_{i}^{\QCD}F_{i}^{\QCD}(\mbb|\vec{\alpha}_i) + \muZ N_{i}^{\zbb}(\vec{\theta}) F_{i}^{\zbb}(\mbb|\vec{\theta}_S)  \\
                                 & + (\muVBF N_{i}^{\qqH}(\vec{\theta}) + N_{i}^{\ggF}(\vec{\theta})) F_{i}^{\hbb}(\mbb|\vec{\theta}_S).
\end{aligned}
\end{equation}

The fit is performed with two unconstrained parameters: the signal strength modifier for the VBF process (\muVBF) and \muZ.
The measurement yields
\begin{align*}
        \muVBF & = 1.01^{+0.39}_{-0.24}\syst\pm0.39\stat,\quad\text{and} \\
        \muZ   & = 0.96\pm 0.21\syst\pm 0.23\stat.
\end{align*}

\begin{figure} [htb!]
        \centering
        \includegraphics[width=.45\linewidth]{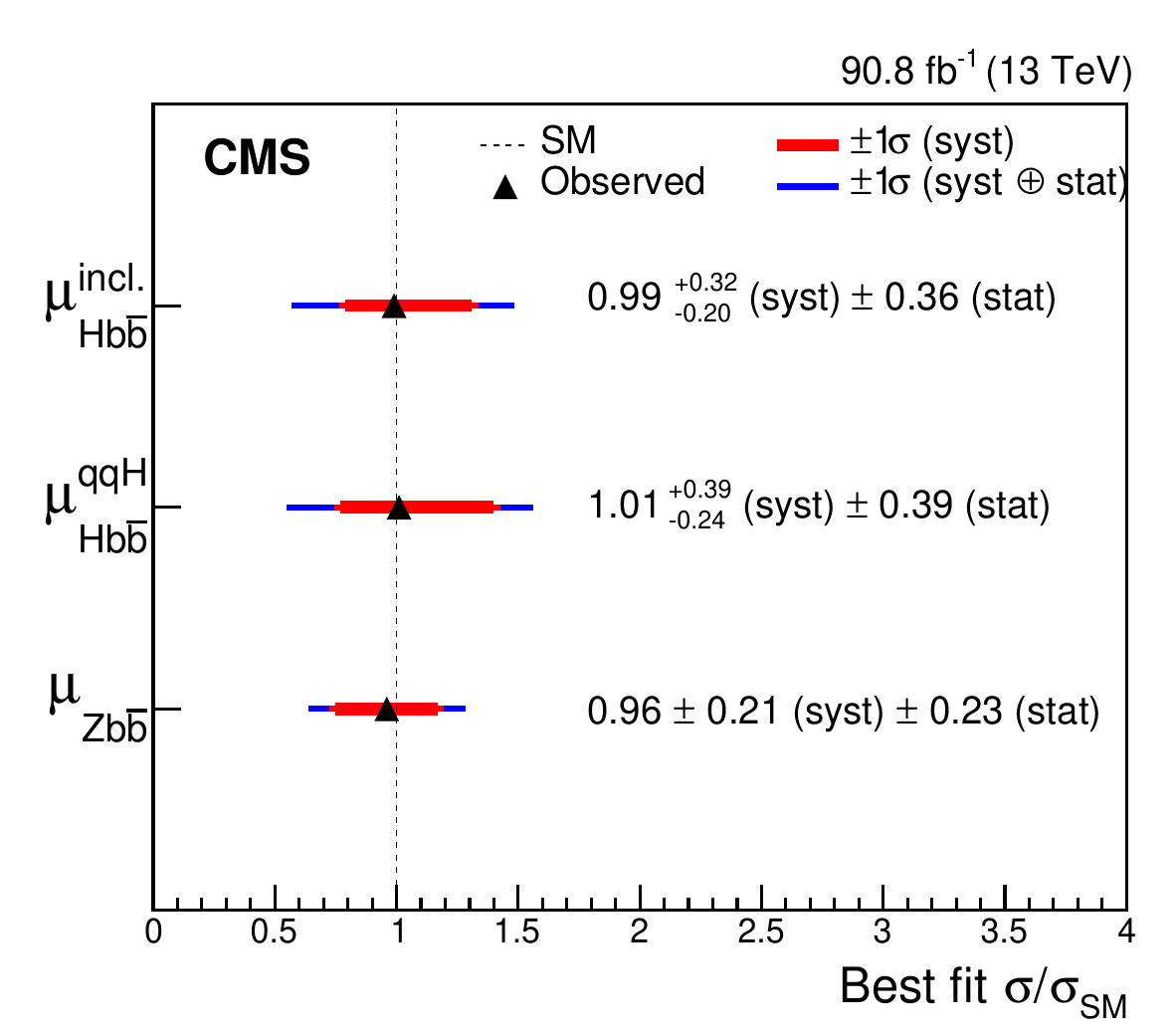}
        \caption{
                The best fit values of the signal strength modifier for the different processes.
                The horizontal bars in blue and red colors represent the ${\pm}1\,\sigma$ total uncertainty and its systematic component.
                The vertical dashed line shows the standard model prediction.
        }
        \label{fig:sig_strength}
\end{figure}

The VBF signal is observed with a significance of 2.4\,$\sigma$.
The expected significance is 2.7\,$\sigma$.
The best fit values of the signal strength modifiers for the different processes are also shown in Fig.~\ref{fig:sig_strength}.

\subsection{Independent measurement of VBF and \texorpdfstring{\ggF}{ggF} production}

Independent measurement of the VBF and \ggF production rates is performed by fitting the \mbb spectra with three unconstrained parameters, \muZ, \muVBF, and \muggF, with the last one being the signal strength for the \ggF process.
In this measurement, the \mbb spectrum in the $i$th category is fitted with the function
\begin{equation}
\begin{aligned}
        F_{i}(\mbb|\vec{\theta}) & =  N_{i}^{\QCD}F_{i}^{\QCD}(\mbb|\vec{\alpha}_i) + \muZ N_{i}^{\zbb}(\vec{\theta}) F_{i}^{\zbb}(\mbb|\vec{\theta}_S)  \\
                                 & + (\muVBF N_{i}^{\qqH}(\vec{\theta}) + \muggF N_{i}^{\ggF}(\vec{\theta})) F_{i}^{\hbb}(\mbb|\vec{\theta}_S).
\end{aligned}
\end{equation}

The fit yields
\begin{equation*}
\begin{aligned}
        \muVBF & = 1.59^{+0.63}_{-0.72}\syst\pm0.54\stat,             \\
        \muggF & = -2.7^{+5.6}_{-2.1}\syst\pm3.5\stat,\quad\text{and} \\
        \muZ   & = 0.94\pm 0.20\syst\pm 0.26\stat.
\end{aligned}
\end{equation*}

\begin{figure} [htb!]
        \centering
        \includegraphics[width=.45\linewidth]{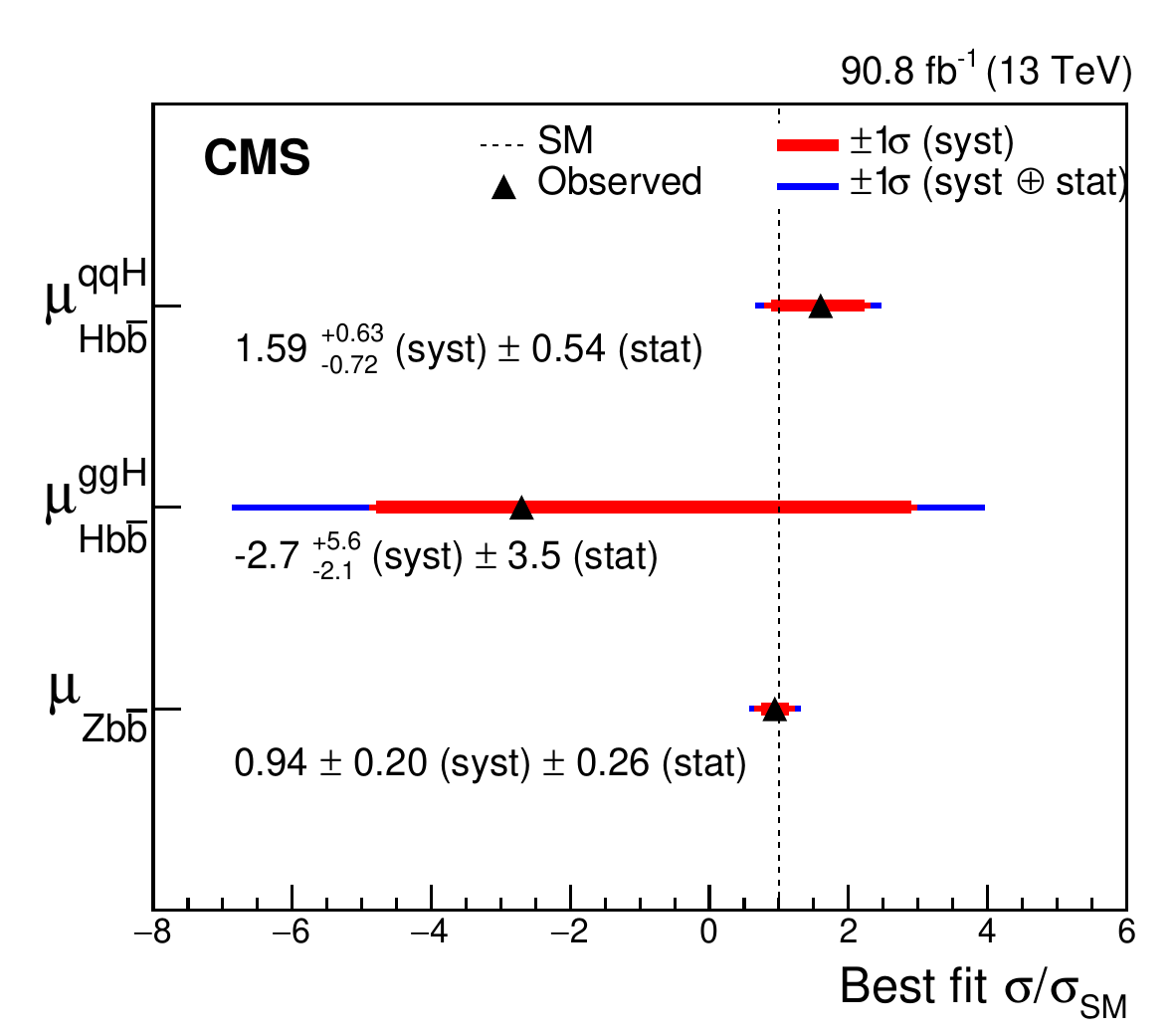}
        \includegraphics[width=.45\linewidth]{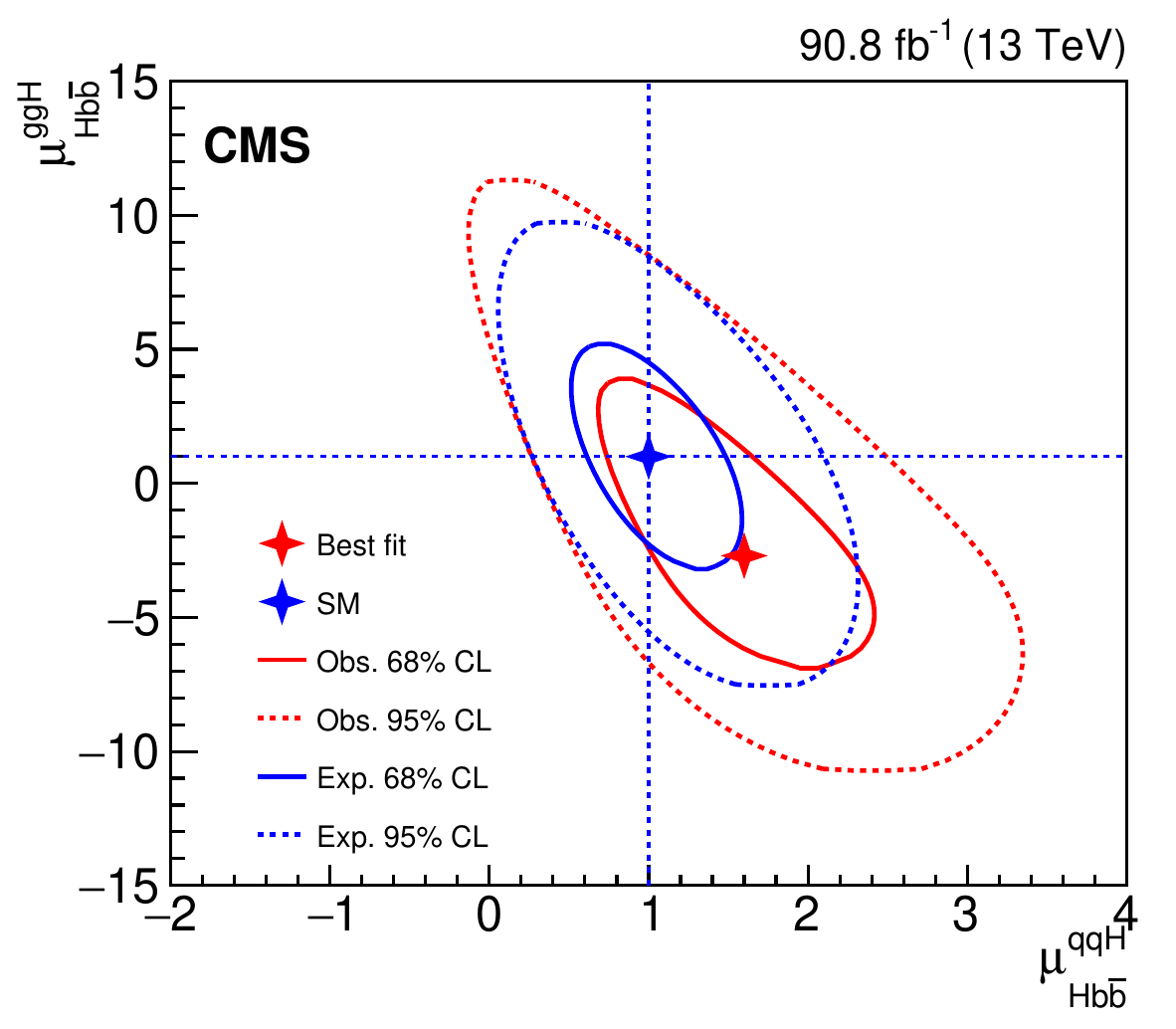}
        \caption{
                The best fit values of the signal strength modifier for the different processes, the horizontal bars in blue and red colors represent the $\pm1\,\sigma$ total uncertainty and its systematic component and the vertical dashed line shows the SM prediction (left).
                The two-dimensional likelihood scan of \muVBF and \muggF, the red (blue) solid and dashed lines correspond to the observed (expected) 68 and 95\% \CL contours in the $(\muVBF, \muggF)$ plane (right).
                The SM predicted and observed best fit values are indicated by the blue and red crosses.
        }
        \label{fig:3_POI}
\end{figure}

The best fit values of the signal strength modifiers for the different processes are shown in Fig.~\ref{fig:3_POI} (left).
The negative best fit value for the signal strength modifier of the \ggF process, \ie, \muggF, is caused by deficit of the data compared to the expected yields from the background processes only for \mbb between 100 and 150\GeV, observed in some of the event categories targeting the \ggF or \zjets processes.
A two-dimensional likelihood scan of \muVBF and \muggF presented in Fig.~\ref{fig:3_POI} (right), shows that the measured value of \muggF is consistent with the SM prediction within 1\,$\sigma$.
Because of the degeneracy of the \mbb distribution shapes in the VBF and \ggF processes and the nonnegligible contribution from the \ggF process in the categories targeting VBF production, \muVBF and \muggF exhibit strong anticorrelation with a linear correlation coefficient of $-0.63$.
As a consequence, a downward shift in the measured value of \muggF causes an upward shift in the measured value of \muVBF.

\section{Summary}
\label{sec:summary}

A measurement of the Higgs boson (\PH) production via vector boson fusion (VBF) process and its decay to a bottom quark-antiquark pair (\bbbar) was performed on proton-proton collision data sets collected by the CMS experiment at $\sqrt{s}=13\TeV$ corresponding to a total integrated luminosity of 90.8\fbinv.
The analysis employs boosted decision trees (BDTs) to discriminate the signal against major background processes: QCD-induced multijet production and \zjets events.
The BDTs exploit kinematic properties of the VBF jets, information of the \PQb-tagged jets assigned to the \Hbb decay, and global event shape variables.
Based on the BDT response, multiple event categories are introduced, targeting the VBF, gluon-gluon fusion (\ggF), and \zjets processes to achieve a maximum sensitivity for the signal.
While the VBF categories have the highest signal-to-background ratio, the \zjets categories constrain the largest resonant background.
The \ggF categories enhance the sensitivity to the inclusive production of the Higgs boson in association with two jets.

The VBF Higgs boson production rate has been measured in its decay to bottom quark-antiquark pairs with the \ggF contribution constrained within the theoretical and experimental uncertainties to the standard model prediction.
The signal strength of the VBF Higgs production, followed by the \Hbb decay, defined as the rate of the signal process relative to the value predicted in the standard model, is measured to be, ${\muVBF=1.01^{+0.55}_{-0.46}}$.
The signal was observed with a significance of 2.4 standard deviations, compared to the expected significance of 2.7 standard deviations.
In addition, inclusive Higgs boson production in association with two jets, followed by \Hbb decay, was measured by treating the \ggF contribution as part of the signal.
The inclusive signal strength was measured to be ${\muHbb=0.99^{+0.48}_{-0.41}}$, corresponding to an observed (expected) significance of 2.6 (2.9) standard deviations.
The measurements are consistent within uncertainties with the prediction from the standard model.

\begin{acknowledgments}
We congratulate our colleagues in the CERN accelerator departments for the excellent performance of the LHC and thank the technical and administrative staffs at CERN and at other CMS institutes for their contributions to the success of the CMS effort. In addition, we gratefully acknowledge the computing centers and personnel of the Worldwide LHC Computing Grid and other centers for delivering so effectively the computing infrastructure essential to our analyses.         Finally, we acknowledge the enduring support for the construction and operation of the LHC, the CMS detector, and the supporting computing infrastructure provided by the following funding agencies: SC (Armenia), BMBWF and FWF (Austria); FNRS and FWO (Belgium); CNPq, CAPES, FAPERJ, FAPERGS, and FAPESP (Brazil); MES and BNSF (Bulgaria); CERN; CAS, MoST, and NSFC (China); MINCIENCIAS (Colombia); MSES and CSF (Croatia); RIF (Cyprus); SENESCYT (Ecuador); MoER, ERC PUT and ERDF (Estonia); Academy of Finland, MEC, and HIP (Finland); CEA and CNRS/IN2P3 (France); BMBF, DFG, and HGF (Germany); GSRI (Greece); NKFIH (Hungary); DAE and DST (India); IPM (Iran); SFI (Ireland); INFN (Italy); MSIP and NRF (Republic of Korea); MES (Latvia); LAS (Lithuania); MOE and UM (Malaysia); BUAP, CINVESTAV, CONACYT, LNS, SEP, and UASLP-FAI (Mexico); MOS (Montenegro); MBIE (New Zealand); PAEC (Pakistan); MES and NSC (Poland); FCT (Portugal); MESTD (Serbia); MCIN/AEI and PCTI (Spain); MOSTR (Sri Lanka); Swiss Funding Agencies (Switzerland); MST (Taipei); MHESI and NSTDA (Thailand); TUBITAK and TENMAK (Turkey); NASU (Ukraine); STFC (United Kingdom); DOE and NSF (USA).

\hyphenation{Rachada-pisek} Individuals have received support from the Marie-Curie program and the European Research Council and Horizon 2020 Grant, contract Nos.\ 675440, 724704, 752730, 758316, 765710, 824093, 884104, and COST Action CA16108 (European Union); the Leventis Foundation; the Alfred P.\ Sloan Foundation; the Alexander von Humboldt Foundation; the Science Committee, project no. 22rl-037 (Armenia); the Belgian Federal Science Policy Office; the Fonds pour la Formation \`a la Recherche dans l'Industrie et dans l'Agriculture (FRIA-Belgium); the Agentschap voor Innovatie door Wetenschap en Technologie (IWT-Belgium); the F.R.S.-FNRS and FWO (Belgium) under the ``Excellence of Science -- EOS" -- be.h project n.\ 30820817; the Beijing Municipal Science \& Technology Commission, No. Z191100007219010; the Ministry of Education, Youth and Sports (MEYS) of the Czech Republic; the Shota Rustaveli National Science Foundation, grant FR-22-985 (Georgia); the Deutsche Forschungsgemeinschaft (DFG), under Germany's Excellence Strategy -- EXC 2121 ``Quantum Universe" -- 390833306, and under project number 400140256 - GRK2497; the Hellenic Foundation for Research and Innovation (HFRI), Project Number 2288 (Greece); the Hungarian Academy of Sciences, the New National Excellence Program - \'UNKP, the NKFIH research grants K 124845, K 124850, K 128713, K 128786, K 129058, K 131991, K 133046, K 138136, K 143460, K 143477, 2020-2.2.1-ED-2021-00181, and TKP2021-NKTA-64 (Hungary); the Council of Science and Industrial Research, India; the Latvian Council of Science; the Ministry of Education and Science, project no. 2022/WK/14, and the National Science Center, contracts Opus 2021/41/B/ST2/01369 and 2021/43/B/ST2/01552 (Poland); the Funda\c{c}\~ao para a Ci\^encia e a Tecnologia, grant CEECIND/01334/2018 (Portugal); the National Priorities Research Program by Qatar National Research Fund; MCIN/AEI/10.13039/501100011033, ERDF ``a way of making Europe", and the Programa Estatal de Fomento de la Investigaci{\'o}n Cient{\'i}fica y T{\'e}cnica de Excelencia Mar\'{\i}a de Maeztu, grant MDM-2017-0765 and Programa Severo Ochoa del Principado de Asturias (Spain); the Chulalongkorn Academic into Its 2nd Century Project Advancement Project, and the National Science, Research and Innovation Fund via the Program Management Unit for Human Resources \& Institutional Development, Research and Innovation, grant B05F650021 (Thailand); the Kavli Foundation; the Nvidia Corporation; the SuperMicro Corporation; the Welch Foundation, contract C-1845; and the Weston Havens Foundation (USA).
\end{acknowledgments}

\bibliography{auto_generated}
\cleardoublepage \appendix\section{The CMS Collaboration \label{app:collab}}\begin{sloppypar}\hyphenpenalty=5000\widowpenalty=500\clubpenalty=5000\input{HIG-22-009-public-authorlist.tex}\end{sloppypar}
\end{document}

%% file: HIG-22-009-public-authorlist.tex
\cmsinstitute{Yerevan Physics Institute, Yerevan, Armenia}
{\tolerance=6000
A.~Hayrapetyan, A.~Tumasyan\cmsAuthorMark{1}\cmsorcid{0009-0000-0684-6742}
\par}
\cmsinstitute{Institut f\"{u}r Hochenergiephysik, Vienna, Austria}
{\tolerance=6000
W.~Adam\cmsorcid{0000-0001-9099-4341}, J.W.~Andrejkovic, T.~Bergauer\cmsorcid{0000-0002-5786-0293}, S.~Chatterjee\cmsorcid{0000-0003-2660-0349}, K.~Damanakis\cmsorcid{0000-0001-5389-2872}, M.~Dragicevic\cmsorcid{0000-0003-1967-6783}, A.~Escalante~Del~Valle\cmsorcid{0000-0002-9702-6359}, P.S.~Hussain\cmsorcid{0000-0002-4825-5278}, M.~Jeitler\cmsAuthorMark{2}\cmsorcid{0000-0002-5141-9560}, N.~Krammer\cmsorcid{0000-0002-0548-0985}, L.~Lechner\cmsorcid{0000-0002-3065-1141}, D.~Liko\cmsorcid{0000-0002-3380-473X}, I.~Mikulec\cmsorcid{0000-0003-0385-2746}, J.~Schieck\cmsAuthorMark{2}\cmsorcid{0000-0002-1058-8093}, R.~Sch\"{o}fbeck\cmsorcid{0000-0002-2332-8784}, D.~Schwarz\cmsorcid{0000-0002-3821-7331}, M.~Sonawane\cmsorcid{0000-0003-0510-7010}, S.~Templ\cmsorcid{0000-0003-3137-5692}, W.~Waltenberger\cmsorcid{0000-0002-6215-7228}, C.-E.~Wulz\cmsAuthorMark{2}\cmsorcid{0000-0001-9226-5812}
% ...
\par}
\vskip\cmsinstskip
\dag:~Deceased\\
$^{1}$Also at Yerevan State University, Yerevan, Armenia\\
$^{2}$Also at TU Wien, Vienna, Austria\\
% ...
